\documentclass[dvipsnames]{article}
\pdfoutput=1
\usepackage{microtype}
\usepackage{graphicx}
 \usepackage{float}
\usepackage{lipsum}% http://ctan.org/pkg/lipsum
\usepackage{graphicx}
\usepackage{booktabs} % for professional tables
\usepackage{hyperref}

\usepackage[accepted]{mlsys2020}
\graphicspath{images/}
 \usepackage{booktabs}
\usepackage{caption}
\usepackage{subcaption}
\usepackage{pgffor}
\usepackage{tikz,subcaption}
\usetikzlibrary{calc, shapes, backgrounds}
\usetikzlibrary{chains, fit, positioning,calc}
\usepackage{amsmath, amssymb}
\usepackage{environ}

\NewEnviron{FitToWidth}[1][\columnwidth]{%
  \makebox[\columnwidth]{\resizebox{#1}{!}{\BODY}}%
}
\usepackage{changepage}
\usetikzlibrary{shapes,arrows}

\mlsystitlerunning{PACSET (Packed serialized trees): Reducing Inference Latency for Tree Ensemble Deployments}
\graphicspath{{images//}{images/plots_bin_depth}}

\begin{document}

\twocolumn[
\mlsystitle{PACSET (Packed serialized trees): Reducing Inference Latency for Tree ensemble Deployment}

\mlsyssetsymbol{equal}{*}

\begin{mlsysauthorlist}
\mlsysauthor{Meghana Madhyastha}{csjhu}
\mlsysauthor{Kunal Lillaney}{amazon}
\mlsysauthor{James Browne}{westpoint}
\mlsysauthor{Joshua Vogelstein}{bmejhu}
\mlsysauthor{Randal Burns}{csjhu}

\end{mlsysauthorlist}

\mlsysaffiliation{csjhu}{Department of Computer Science, Johns Hopkins University, Baltimore, USA}
\mlsysaffiliation{bmejhu}{Department of Biomedical Engineering, Johns Hopkins University, Baltimore, USA}
\mlsysaffiliation{amazon}{Amazon, Seattle, USA}
\mlsysaffiliation{westpoint}{Department of Computer Science, United States Military Academy, West Point}

\mlsyscorrespondingauthor{Meghana Madhyastha}{mmadhya1@jhu.edu}

\mlsyskeywords{Machine Learning, MLSys}

\vskip 0.3in

\begin{abstract}
 % Abstract needs to be concise.
We present methods to serialize and deserialize tree ensembles that optimize inference latency 
when models are not already loaded into memory. This arises whenever models are larger
than memory, but also systematically when models are 
deployed on low-resource devices, such as in the Internet of Things, or run as Web micro-services
where resources are allocated on demand.
Our packed serialized trees (PACSET) encode reference locality in the layout of a tree ensemble
using principles from external memory algorithms. The layout interleaves 
correlated nodes across multiple trees, uses leaf cardinality to collocate the nodes on the most 
popular paths and is optimized for the I/O blocksize. The result is that each I/O yields
a higher fraction of useful data, leading to a 2-6 times reduction in classification latency for 
interactive workloads. 
\end{abstract}
]
\printAffiliationsAndNotice{} % otherwise use the standard text.
\section{Introduction}
% Big problem in science 
With the widespread deployment of machine learning, researchers have
turned their focus toward performance and reliablity of production systems.
Example application areas include as personal digital assistants \cite{stasior2017zero}, video surveillance \cite{ananthanarayanan2017real}, and  
directed advertising \cite{cai2017real},  % citations from clipper and elsewhere.  pick top papers
Many frameworks deploy services in the cloud \cite{olston2017tensorflow} \cite{moritz2018ray}
and focus on issues, such as multi-tenancy, model distribution, and scalability \cite{8360337}.
Inference latency becomes a critical metric when deploying machine learning as a service (MLaaS) \cite{inproceedings} % 
integrated into Web and mobile applications. Many applications rely on real-time prediction \cite{ananthanarayanan2017real} to deliver
model outputs to apps, dashboards, and systems, examples include real-time bidding systems, financial trading, and
predictive maintenance. Google provides extensive guidance on how to deploy systems for real-time inference that include
reducing model size, fast I/O systems, caching, and accelerators (GPUs and TPUs) \cite{googlearticle}. %https://cloud.google.com/solutions/machine-learning/minimizing-predictive-serving-latency-in-machine-learning
The Clipper system \cite{crankshaw2017clipper}  
 builds middleware on top of existing machine learning frameworks, 
such as scikit-learn and tensorflow, that reduces latency, boosts throughput, and increase accuracy.

% Narrower problem
Despite real time requirements, most machine learning algorithms and frameworks continued to be designed for batch throughput. 
This problem is particularly acute for ensemble learning, including gradient-boosted trees and random forests.
High-performance tree ensembles, XGBoost~\cite{chen2016xgboost} and LightGBM~\cite{ke2017lightgbm}, use many techniques to achieve parallelism
and throughput, including model pruning, sampling, and discretization. However, they measure prediction performance as the
aggregate throughput of inference using batches of large size; papers do not even report the latency of a single classification.
This reflects the co-design of the system with its intended and most frequent usage, offline prediction that scales with large data size. Performance matters, but is not measured by latency. For example, Kaggle competitions have kernel limits on compute time and memory consumption, but under those limits are evaluated based on model performance. This setting values throughput, which allows for more complex models under the resource budget.

% Yet narrower paper gap 
At the same time, tree ensembles are particularly attractive for distributed deployment because inference requires a minimal amount of computation, which makes hardware inexpensive and energy consumption low when compared with deep learning \cite{lowenergyrf}. %https://ieeexplore.ieee.org/document/8957193
Recent efforts to put deep learning on edge devices have revealed that models are often too large and that compute
resources are inadequate, %https://dl.acm.org/doi/pdf/10.1145/3229556.3229562?download=true
leading to a hybrid design that splits computing across the edge and cloud servers \cite{8270639}. %https://www.researchgate.net/publication/322728184_Learning_IoT_in_Edge_Deep_Learning_for_the_Internet_of_Things_with_Edge_Computing % https://ieeexplore.ieee.org/document/8270639
Model compression is another approach to fit neural networks onto edge devices \cite{lowenergyrf}. %%https://ieeexplore.ieee.org/document/8957193
Tree ensembles require the same level of attention to latency and edge deployments. Tree ensembles are the preferred method for
many problems \cite{fernandez2014we} %http://jmlr.org/papers/volume15/delgado14a/delgado14a.pdf % https://dl.acm.org/doi/abs/10.5555/2946645.3007063,
particularly when there is limited input data or categorical features.  

% Summary our approach--our results
We present techniques that reorganize tree ensembles to reduce latency when models are located on storage or in external memory.  %external in solid state storage or in a key-value store such as Redis
These techniques also eliminate the need to keep models small enough to fit in RAM.
Packed Serialization Trees (PACSET) reorganize the layout of trees to minimize I/O by performing {\em selective access}: only the parts of the model needed for inference are loaded into memory. PACSET serializes a forest into a sequence of bytes that can be stored as a file or streamed over networks. 
To minimize I/O, PACSET places tree-nodes that are accessed together into the same storage block, 1KB to 256KB depending on device
properties. Inference in a forest accesses a single path in each tree. This is a small amount of data. 
Thus, grouping data requires PACSET to identify locality that span multiple trees. PACSET interleaves the
top levels of multiple trees, uses leaf cardinality to cluster nodes that are in popular paths, and optimizes the layout for a given I/O block size. Taken together, layout optimizations reduce latency by 2-6 times for large datasets.
PACSET produces the same output as unoptimized trees, avoiding the runtime/performance tradeoffs of model pruning or 
discretization.

Our final submission will include a reproducible artifact as defined by the ACM Artifcat Review and Badging policy. Our release will include open-source repostiories under a permissive license and containers that include the exact build used to run experiments. The experimental hardware can replicated exactly on the cloud, with the exception of embedded systems.

\section{Related Work} %Move related work up front
%The past few years has seen a variety of work focused on scalable ensemble algorithms and implementations. In %this section, we briefly describe some of the recent literature  to put our contributions in perspective.

Several systems implement scalable and high-throughput gradient boosted trees. %One of the most popular projects in this space is XGBoost.
XGBoost~\cite{chen2016xgboost} is an open source library for the scalable training and inference of gradient boosted trees using techniques such as cache access patterns, data compression and sharding. Sparsity aware training and a novel sampling technique reduces the memory footprint of XGBoost on large datasets \cite{alafate2019faster}. LightGBM \cite{ke2017lightgbm} is a library for gradient boosted trees that implements sampling techniques to alleviate the bottlenecks that arise from the $O(n)$ scan and split at each node during training. The optimizations in XGBoost as well as LightGBM serve primarily to reduce training and inference time for batches of observations. Treelite~\cite{cho2018treelite} generates C source from XGBoost models and compiles models into a shared library. Compilation encodes branch prediction and model quantization to enable faster and memory efficient inference.

There is also research into optimizing random forests.
A hybrid BFS-DFS layout of nodes increases training performance, but also increases memory footprint \cite{anghel2019breadth}. 
%Consequently the size of main memory required for training also grows for deep tree ensembles.   
Pruning random forests reduces model size for inference \cite{painsky2018lossless}. Hummingbird \cite{hummingbird} converts models to tensor arithmetic so that inference can be performed by deep learning frameworks.  Hummingbird is not well suited to deep tree ensembles because the tensor size grows exponentially with tree depth.
A cache-aware tree layout results in efficient inference, but forest models must fit in memory \cite{anonymous}. The RAPIDS Library \cite{rapids} optimizes tree layouts for inference on GPUs. Tree ensembles must fit in GPU memory. 

Tree ensembles are far more energy efficient and require much less memory and compute than deep learning methods and, thus, better suited to embedded systems. Race logic implementations of ensemble learners show how a programmable accelerator can be used for machine learning implementations  \cite{tzimpragos2019boosted}. Zhao {\em et al.} \cite{zhao2019rfacc} implement a ReRAM based accelerator that acceleratess random forest training. Bonsai \cite{kumar2017resource} learns a single, shallow, sparse tree in a low dimensional space that the original data is projected onto to support low-resource IoT devices.

Serverless computing supports the on-demand allocation of compute to implement Web microservices. Most research focuses on model training in a serverless framework, rather than the model inference and model serving. Pywren uses AWS lambda to train machine learning models at massive scale \cite{DBLP:journals/corr/JonasVSR17} in map-reduce framework. Cirrus \cite{10.1145/3357223.3362711} is a microservice framework that uses a distributed data store for model representations. It only supports  a subset of algorithms that use stochastic gradient descent.  
%such as Sparse Logistic Regression, Latent Dirichlet Allocation, Softmax and Collaborative Filtering. These works on 
%ML in the cloud focus on model training in a serverless framework and not model inference or model serving. 
Ishakian et al.~\cite{8360337} evaluate the suitability of a serverless computing for inference on large neural networks.

Out-of-core machine learning has been implemented in several settings. knor \cite{mhembere17knor} is a clustering library with external memory support. The Vaex library \cite{vaex:2009} supports visualization and analysis of large tabular datasets stored in Pandas-like dataframes.
There are memory efficient data structures for out-of-core machine learning \cite{eadsmemory} . A general shortcoming of such libaries is that their optimizations are geared towards \textit{\textbf{data}} that does not fit into memory. They require the entire model to be loaded into memory.

%Figure 1: just a)

\section{Design of PACSET Serialization}

Our design minimizes latency anytime the model is stored in external memory and not already loaded into main memory.  We consider three deployment scenarios: large models that exceed the RAM of compute nodes  (\S~\ref{sec:pacsetlarge}), cloud microservices (\S~\ref{sec:pacsetaas}), and edge devices with limited memory resources (\S~\ref{sec:pacsetemb}).
Microservices allocate a compute resource on demand and load data from a cloud storage service, e.g. an in-memory object store such as AWS Elasticache Redis. Edge devices use flash to store data because it is cheaper than RAM and
persistent. Edge devices have a small amount of RAM (internal memory) to be used at runtime. 

We derive the hyperparameters of PACSET from the properties of the I/O system and the dataset.
Solid state storage devices perform best using an I/O size that is between 16-256K.
Although the minimum I/O size and device block size may be smaller, e.g. 4 KB, SSDs have parallel channels so that sequential 
and aligned I/O across all channels benefits performance. 
Tree ensembles can be large and complex, with high-dimensional inputs, 
many output classes, and millions or billions of samples. The resulting forests have 
hundreds or thousands of trees that are often 12 or more levels deep. 
Random forests often train until leaf nodes are pure, i.e. encode a single class.
Efficient gradient-boosted forests, which prefer many shallow trees, recommend
trees that are ten levels deep \cite{chen2016xgboost}. 

We reduce latency by minimizing the number of block I/Os needed to perform a single inference in a forest.
An inference accesses a single path in each tree. Most frameworks \cite{chen2016xgboost} \cite{ke2017lightgbm} \cite{pedregosa2011scikit} 
pack multiple trees into a single file, but serialize each tree independently. There may be some haphazard I/O savings, particularly when trees are smaller than blocks. Packed serialized trees (PACSET) identifies correlation between the nodes and paths across trees so that they appear in the same block. Some correlations are systematic; the root node of every tree is accessed during every inference, so placing all root nodes in a single block is helpful. The more interesting relationships are statistical. Within trees, we group the nodes in the most popular paths based on leaf cardinality. Further, our packing layouts are block size aware so that each block starts with a high cardinality node.

We now describe the main components of our system at a high level. Broadly, there are three phases involved in the process: training, packing and inference. 
% \fboxrule=1.5pt%border thickness
% \begin{figure}
%   \begin{subfigure}[b]{\columnwidth}
%       \input{mlsys2020style/tikz_figures/blockpack}
%         %\caption{PACSET Pack}
%         \label{subfig:blockpack}
%     \end{subfigure}
    
%     \begin{subfigure}[b]{\columnwidth}
%       \input{mlsys2020style/tikz_figures/blockinference}
%         %\caption{PACSET Inference}
%         \label{subfig:blockinference}
%   \end{subfigure}
 
%     \caption{System overview of PACSET shows packing phase (top) and inference phase (bottom).}
      
%   \label{fig:block}
% \end{figure}

% \fboxrule=1.5pt%border thickness
% \begin{figure}
%   \begin{subfigure}{\columnwidth}
%       \includegraphics[width = \columnwidth]{mlsys2020style/images/f2.png}
         
%         \label{subfig:blockpack}
%     \end{subfigure}
    
%     \begin{subfigure}{\columnwidth}
%       \includegraphics[width = \columnwidth]{mlsys2020style/images/f1.png}
        
%         \label{subfig:blockinference}
%   \end{subfigure}
 
%     \caption{System overview of PACSET shows packing phase (top) and inference phase (bottom).}
      
%   \label{fig:block}
% \end{figure}

  \begin{figure}
      \includegraphics[width = \columnwidth]{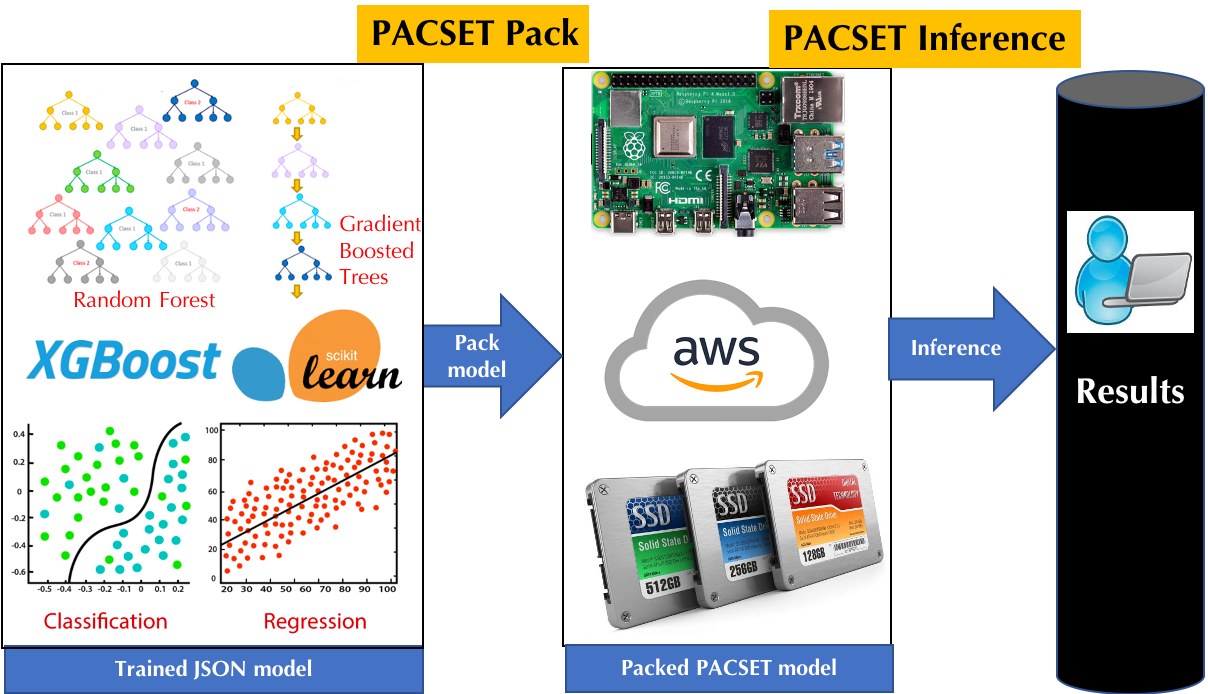}
         \caption{Overview of PACSET packing and inference}
        \label{subfig:blockpack}
 
    \end{figure}

\vspace{-10pt}

\begin{enumerate}
    \addtolength{\itemsep}{-6pt}
    \item \textbf{Training:} PACSET takes as its input a trained tree ensemble. Currently, models trained via XGBoost and Scikit-Learn are supported. 
    \item \textbf{Packing:} The trained model is read and packed (Figure \ref{subfig:blockpack}). \textit{Packing} refers to the organization of the nodes of the tree ensemble in a linear array in which the placement of the nodes is determined by the I/O optimized layouts described in detail in \S\ref{packingalgo}. The packed model is then stored in external memory, either to flash storage such as an SSD (\S\ref{sec:pacsetlarge}), a key value store such as Redis (\S\ref{sec:pacsetaas}), or a microSD card (\S\ref{sec:pacsetemb}).
    \item \textbf{Inference:} PACSET performs inference directly on the model stored in external memory without having to load the entire model into main memory.  PACSET memory maps (mmap) the file on storage  or reads individual nodes from a key-value store when performing inference from microservices in the cloud.
\end{enumerate}

\vspace{-15pt}

\section{Packing Algorithms}\label{packingalgo}
 
 We describe the methods used to pack forests and the resulting layouts. We first describe the the depth-first and breadth-first implementations of existing systems. We then add PACSET's optimizations for interleaving trees, statistical layout, and block awareness. For input, PACSET requires a forest in a standard format (scikit-learn or XGBoost) that includes leaf-cardinality, i.e.~how many of the training samples route to each leaf. PACSET compute the cardinalities of interior nodes from leaf cardinalities.
 
 %interleaving : RB writes it. Interleaving is inspired by disk striping, figure for just interleaving, four trees, much smaller, four trees different colors, first 12 nodes, interleaved, four colors showing striping
 
 %Show residuals, block bin is better, lets imagine simple example, stripe the top two/three layers, four trees, 
 
 %color each tree, 
 %visualize exactly what is happening
 %trees different sizes
 % 2+ eps, 1+eps, 1-2eps
 % 9, 7, 3, 1, 1
 
 %talk about this in terms of three dimensions of optimizations.
 %Optimizations that are interleaving, weighted and block oriented
 %figure 2 needs reworking
 %need a figure that communicates
 
 %Start with
 %In this section we describe our packing layouts in detail. Figure~\ref{fig:treelayout} shows a tree ensemble consisting of two trees. The branches are labelled with the fraction of nodes travelling along the branch. The cardinalities can be computed by multiplying the cardinality at the head node of the branch with branch labels. The cardinality at the root node represents the total cardinality i.e the total number of observations in the training dataset. The serialization layouts corresponding to this ensemble are also shown; from top to bottom, the BFS, DFS,  and blockwise  layouts respectively. These layouts are described below.
 %MMTODO: "Algorithms" in appendix ?

{\noindent \bf{Breadth First Search}}
(BFS) is the baseline layout used in XGBoost \cite{chen2016xgboost}. Each tree is serialized using a breadth-first traversal of the tree as output by training. Successive trees are output one after another. BFS is the best choice for shallower trees and large batch sizes.

{\noindent \bf{Depth First Search}}
(DFS), as used in scikit-learn \cite{pedregosa2011scikit}, serializes trees with a depth-first traversal. Successive trees are output one after another. DFS is preferred to BFS for deeper trees and smaller batch sizes.

%RB -- pseudocode for weighted-dfs.  Progresssive figures. BFS labelling (color %coding) or weighted labelling (color coding)??
%Present as BFS/DFS and then 3 sections that build.

%\subsubsection{Depth First Search}
%In the depth First Search (DFS) layout, each tree is arranged in a %depth first search order in a linear fashion. DFS traversals %starts at the tree root and moves to one of its children. %Traversal then continues recursively to this child's furthest %descendent before exploring the other child. Thus traversal occurs %in a depth first manner as the entire depth corresponding to a %node is explored before moving on to the next node.
\begin{figure}
\includegraphics[width=\columnwidth]{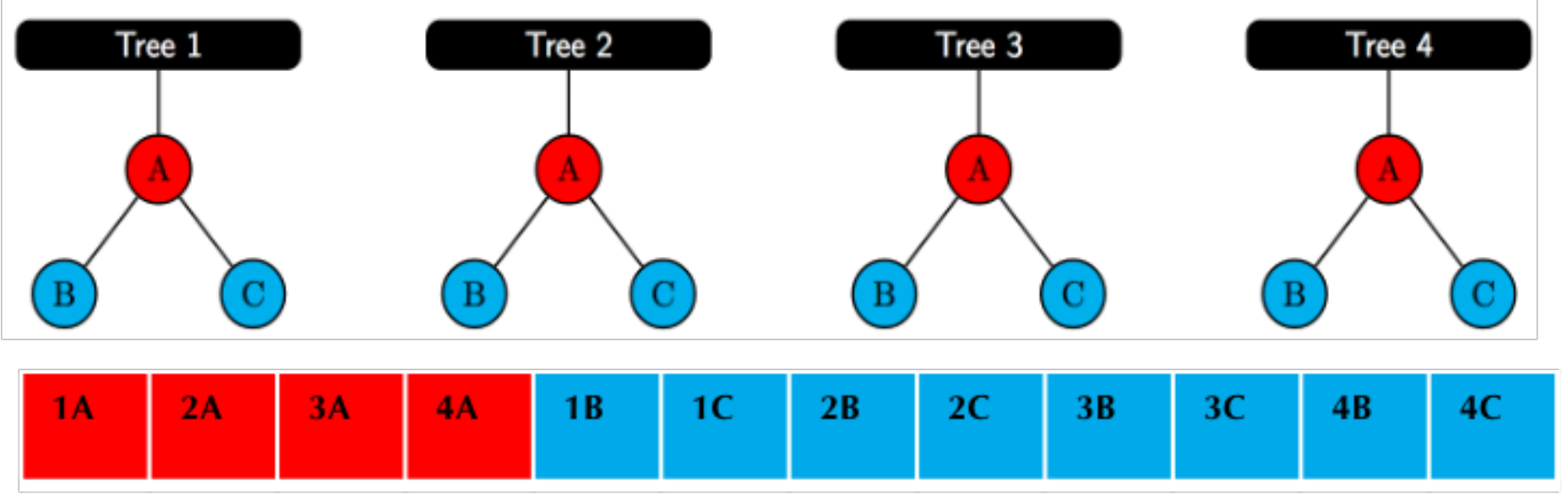}
\caption{The interleaved bin layout stripes the nodes of multiple trees at each level sequentially.}
\label{fig:bin}
\end{figure}

\subsection{Interleaved Bins (BIN)} PACSET's first optimization interleaves or stripes the top levels nodes of multiple trees and packs them into a single I/O block (Figure \ref{fig:bin}). With this layout, loading a single block from external memory fetches the prefix of all paths in the top levels. 
Furthermore, a large fraction of the data within the block will be used. Inference navigates a root to leaf path in each tree. All of the root nodes are evaluated, 50\% of the level 1 nodes, 25\% of the level two nodes, etc.
Two parameters dictate the layout. These are the number of trees per bin and the depth of interleaving. 
We call the remaining sub-trees below the interleaved bin {\em residual} trees or {\em residuals} and they will be packed into separate blocks using a statistical layout. The figure shows an interleaved depth of two. This creates four residual trees for each tree (not shown). A depth of three creates eight residuals.

The best choice of parameters interleaves few levels across many trees. The specific choice depends on the block size and latency of external memory and we will determine best parameters experimentally.  As we interleave more levels of the tree, each level encodes less useful data. Multiple trees increase the amount of memory parallelism. Informally, two to four levels of interleaving work best and we choose as many trees as possible that fit within a block.

Interleaving encodes static and systematic relationships among the nodes in multiple tree, i.e. all root nodes and 50\% of level 1 nodes are accessed.  In the residuals, node accesses are so sparse that interleaving provides no benefit. Instead, we look to encode statistical relationships along high-probability paths within each three.
%In the layouts described above, the trees are packed one after the other. Interleaving, however, packs all the nodes in a given position across all trees before proceeding to the next node. The depth of the BIN indicates the level of interleaving. For example, an interleaved BIN of depth 1 indicates that the root nodes of all the trees in the ensemble are interleaved before proceeding to pack them in one the layouts described above. An interleaved BIN of depth 2 indicates that the root nodes are interleaved followed by the left children of the root nodes of all the trees followed by the right children of the root nodes of all the trees. Then the trees are packed with a layout described above.

%RBTODO cite james note difference of goals between inclusion and density.
%RBTODO in memory versus external memory comparison Table 3.

\begin{figure}
\includegraphics[width=\columnwidth]{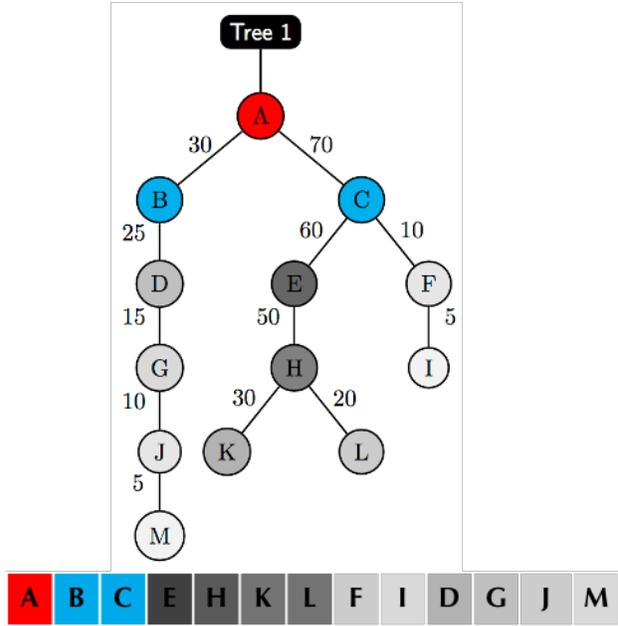}
\caption{Weighted DFS layout of the interior (non-leaf) nodes of a tree.
Each node is annotated with cardinality---number of observations through that node. Grayscale also indicates cardinality. The layout shows that the serialization encodes both cardinality and locality (to the same subtree).}
\label{fig:wdfs}

\vspace{-5pt}

\end{figure}

\subsection{Weighted Depth First Search (WDFS)} 
PACSET serializes residuals by grouping nodes based on the frequency and conditional probability of node accesses.  Starting with root nodes, a weighted depth-first traversal (WDFS) selects the child with the highest cardinality at each node. % This is a depth-first search weighted by cardinality. 
We output nodes that are not part of the interleaved bin.  Figure \ref{fig:wdfs} show a traversal from the root (A) to the highest cardinality child (C). These nodes are in the bin and not output. It then outputs E, H, and then K. Then, traversing back up the tree L and then F.  WDFS ensures that we output the most ``popular'' paths sequentially (EHK), so that a single I/O accesses the entire path with high probability. This path could require two I/Os when the sequence crosses a block boundary. WDFS also groups the entire subtree under a popular node (EHKL), packing subtrees into contiguous regions.  This benefits when subtrees are smaller than a block so that all paths through the popular node requires only a single I/O.

For classification, PACSET serializes the interior nodes of the tree only. Figure  \ref{fig:wdfs} does not show leaf nodes and the cardinalities do not sum. For example, there is a implicit leaf off of node B with cardinality 5 and two leafs off of node L that sum to 20.  For classification, the leaf nodes do not contain information outside of the class and our encoding inlines this information, i.e.~it replaces the pointer to the leaf with the class.  For regression forests, leaf nodes contain counts needed to perform regression and must be stored explicitly. PACSET also packs leaf nodes in this case. We show only classification to simplify figures.
%Figure \ref{fig:pseudo} gives pseudocode for the WDFS algorithm.  

PACSET augments the input forest to compute the access frequency at every node in a tree.
The input forest has leaf cardinalities and class labels. We calculate the cardinality at each node as the subtree sum of leaf cardinalities.
We experimented with performing WDFS from the residual roots, starting at D, E, F, rather than at A. We found this policy to be worse in practice. It favors long narrow paths (DGJM) over short broad paths under a high cardinality node (FI).  This is a minor effect.

% TODO
%RB for caption -- grayscale shows overall cardinality. it does not encode the conditional probability of nodes in the space path
%% MM -- how can cardinalities decrease in single path trees?  Why don't they sum?
%Answer: They will add up if you consider the leaves i.e the remaining observations go to the leaf nodes though in the diagram they just disappear. I did not show the leaves in the figure for brevity.

%With this layout, an access to the top node of a residual will fetch a block that includes all of the most frequently accessed leaves.  Less frequently accessed leaves are deferred to other blocks of storage. The goal is to capture the entire path from the top node of the residual to a leaf in a single block I/O. Depth-first search ensures that we capture entire paths from residual top to leaf. Weighting by cardinality ensures that these are the most frequently accessed paths.

%To encode the statistical relationships between 
%Weighted Depth First Search(WDFS) is a variant of DFS in which direction of exploration is determined by the weights of the children nodes. The child with the higher cardinality is explored first. Here the cardinality of a node can be defined as the number of observations that pass through the node during training. This information is stored and retained during the training phase. When packing, nodes are pushed onto a stack in the order of the cardinalities. 

\begin{figure*}
\includegraphics[width=\textwidth]{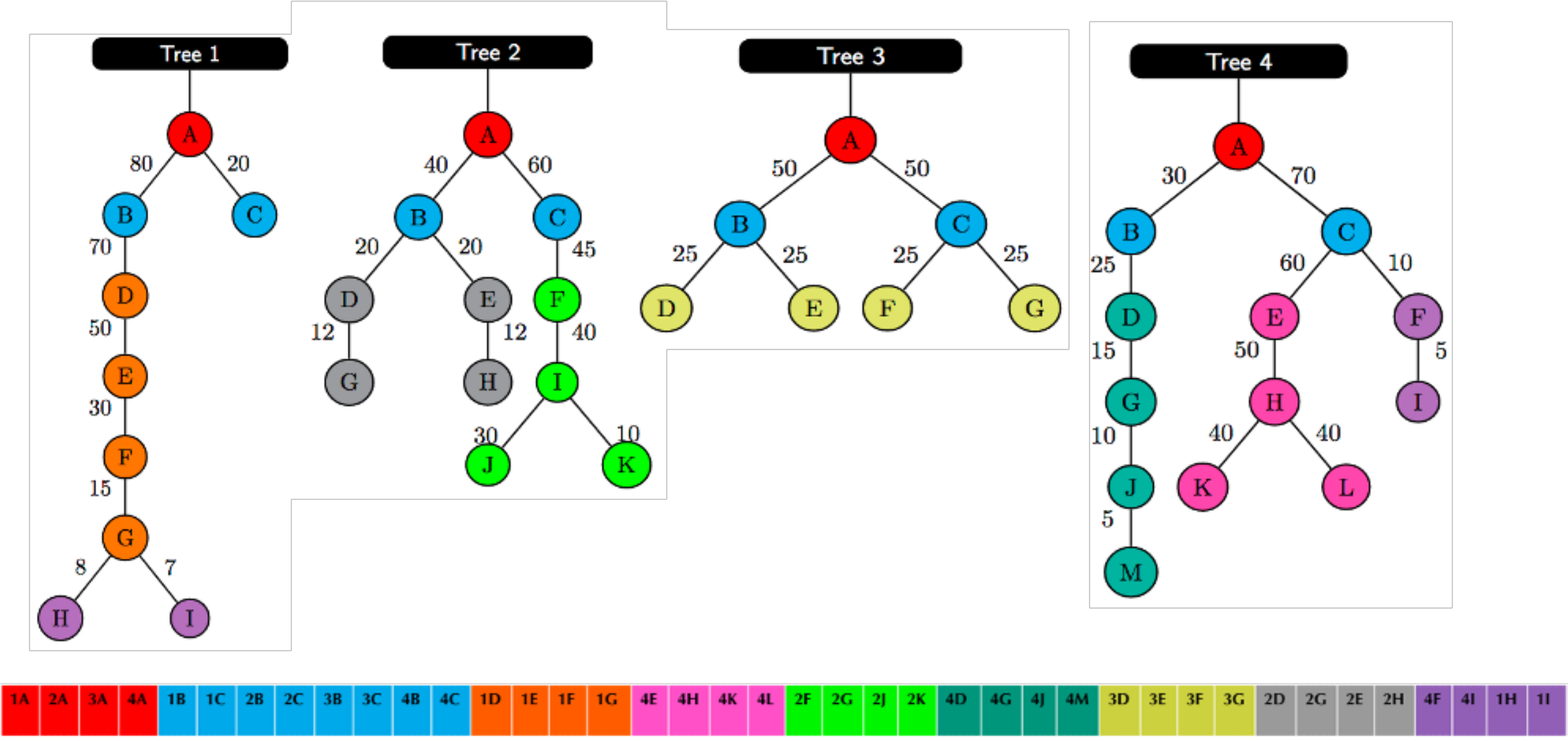}
\caption{Block aligned layout of the interior nodes of a classification forest with a block size of 4.  The layout interleaves the bin (1A-4C) and then groups the highest cardinality paths into blocks. Colors indicate block boundaries for residuals.}
\label{fig:bawdfs}
\end{figure*}

\subsection{Block-aligned Weighted Depth First Search} 

PACSET extends WDFS to make it aware of block boundaries to further reduce I/O.
The goal of an external memory algorithm is to minimize the number of block transfers.  Because blocks are large, 4K-256K corresponding to 128-8192 nodes, a forest packing algorithm aims to capture as many of the most popular paths in a residual in a single block. This is a quite different goal than our prior optimization work on cache-lines \cite{anonymous}, which maximizes the density of useful nodes in small cache lines.  For small cache lines, maximizing density minimized I/O. For PACSET, we are dealing with small numbers. Success is inference in one or twos block I/Os and failure is three or more. The system design reflects this cost function. The goal is to capture as many of the best full paths as possible in a single block.

Block-aligned WDFS halts WDFS at each block boundary and resets the algorithm to start a new block with the highest-cardinality node in the forest.  Figure \ref{fig:bawdfs} shows the resulting layout for a forest using a block size of four nodes.  After the BIN, the algorithm outputs DEFG in tree 1 to fill a block. It then resets and chooses the best subtree of EHKL in tree 4. This defers nodes H and I in tree 1 to a later block that will be filled with nodes from multiple trees all with low cardinality. This is a good result because the traversal of path DEFGH in tree 1 is going to take two block I/Os and node H has no correlation with any other nodes.  By deferring tree 1 node H and I, we prevent these nodes from polluting other blocks. 

Block-aligned WDFS greedily packs the best paths and nodes into blocks and defers low cardinality nodes with no correlation to the end. This has a profound effect of performance.  It maximizes the correlation of nodes within each block by using the block size to reset the algorithm on each boundary.

\section{PACSET Deployment}\label{pacsetdeployment}
We present three deployment scenarios all of which require model data to be accessed from external memory. The scenarios differ in tunable hyperparameters, such as the blocksize and bin depth, that we derive hardware performance properties.  The  packing algorithms are the same. 

\begin{figure*}[t]
     \includegraphics[width=\textwidth]{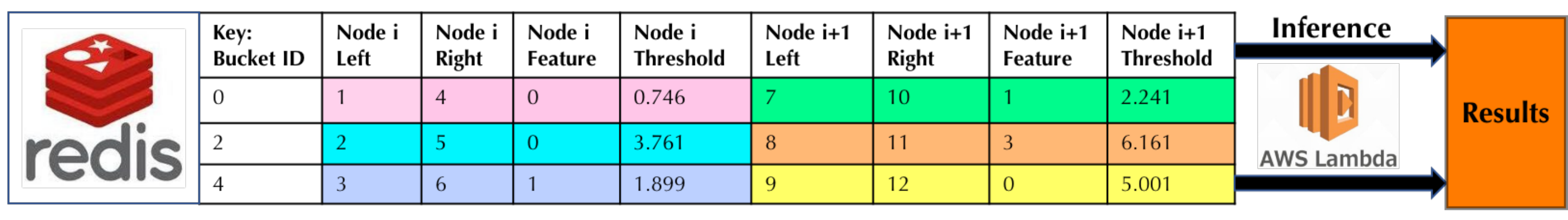}
     
     \vspace{-4pt}
     \caption{This figure shows the overall system architecture of PACSET as a service. The arrangement of nodes in Redis are also shown for a sample of 6 nodes with a bucket capacity of two nodes. Each node is colored differently for visual clarity.}
     \label{fig:lambdadiag}

\end{figure*}

\subsection{Larger than RAM models}\label{sec:pacsetlarge}

When machine learning models grow larger than the capacity of RAM, selective access is needed to perform the learning task.  This can be achieved trivially in forests by loading one tree at a time or trees in small batches. However, this tends to result in high latency.  The standard in both random forests and gradient boosted forests is to assume that the model fits into memory and RAM capacity becomes a de facto limit on model complexity.  PACSET eliminates this barrier, allowing tree ensembles to grow as large as needed for the learning task. 

Tree ensembles are increasingly used for data with high-dimensional feature spaces and billions of samples. Such a ``big data'' application was our original motivation in pursuing this research. Our goal was to classify aging disorders in human connectomes derived from ultra high-resolution MRI \cite{anonymous}. Training a forest on this task until leaves are pure and classification performance stabilized produced a forest of 2048 trees with a total size of 300GB.

Even when tree ensembles are smaller than memory, they are often used in complex machine learning architectures that combine forests with neural networks or stack them into networks of ensembles \cite{zhou17deepforest, feng18mlgbt}. In this case, RAM must be shared and models loaded and unloaded dynamically. PACSET replaces the monolithic load of a model with selective access.

This deployment runs on multicore servers typical of cloud servers or workstations. The model resides on a solid-state storage device.  Performing machine learning inference entails memory mapping the model on disk so that it is paged on demand. In this way, the model is read selectively and lazily; i.e a read triggers a page fault that transfer data from SSD to memory.
Device characteristics determine the number of bytes loaded together, which we call a block. 
The SSDs we use have a 4 KB page and 16 parallel channels. We define a block of 64K so that each read uses the entire device. Each block holds 1024 32 byte tree nodes. Reading any node from block fetches all 1024 nodes together.

%Here, we describe the first and most general deployment scenario, one in which the model simply does not fit into main memory. Here, the model resides on a solid state storage device (SSD). Performing machine learning inference entails memory mapping (mmapping) the model on disk so that it is paged just as it would have been had it been in memory. This way, the model is read selectively and lazily; i.e a fresh read from disk would trigger a page fault and would then be read into memory. Thus, parts of the model are loaded as and when needed. Device characteristics such as the SSD page size determine the number of bytes loaded together. For example, say the SSD page size is 4KB. That means that 4KB are loaded into memory at a time. Lets say the size of a tree node is 32 bytes. Then, (4KB/32bytes)=128 nodes are loaded into memory at once. So upon reading the first node from disk, the next 127 nodes are also read into memory. Our novel blockwise packing algorithm is block size agnostic and packs the nodes together in blocks.

\subsection{PACSET as a service}\label{sec:pacsetaas}

 In serverless computing, applications run as ephemeral functions that are executed as needed. Resource demands are scaled by the cloud service provider. Serverless applications are typically cheaper than applications with dedicated hardware. %Serverless computing is most often used for data parallel applications that are parallelizable across multiple lambda services. For example, map-reduce style operations are amenable to be executed by lambda functions. They are also used as orchestration functions. 
However, a recent study \cite{hellerstein2018serverless} asserts that microservices are neither efficient nor cost effective at machine learning inference. It found that AWS Lambda services are 27 times slower than a dedicated EC2 instance. The current serverless paradigm incurs high I/O latencies  when reading from slow storage, such as S3 buckets. %and more generally, the fact that data and compute are not co-located. Thus, storage latencies involved in lambda communication is one of the main causes of this bottleneck. 

PACSET overcome the I/O limitations of serverless computing by storing models on low-latency key/value services in the cloud and performing selective I/O at a fine granularity. Our ``PACSET-as-a-service'' architecture (Figure~\ref{fig:lambdadiag}) maps a block of tree nodes onto a key/value pair.  The implementation uses AWS Lambda for serverless compute and AWS Elasticache Redis. The best design determined experimentally, uses a very small blocksize of 8 nodes per key/value pair and parallelizes across PACSET bins, creating one Lambda function per bin.

 %Resource constraints in serverless functions hinders scalable deployment capabilities of large machine learning models. This is a deployment application that can benefit from PACSET because PACSET doesn't require the model to be loaded into memory. 
 
 %We coin the term "PACSET as-a-service" for this system and describe the architecture. A schematic diagram of PACSET-as-a-service is given in Figure~\ref{fig:lambdadiag}. Currently, it runs on AWS cloud services but it can be made to support other cloud providers as well. The model is stored in a fast key-value store such as Redis. The Redis server runs on an EC2 instance. AWS lambda functions run the Redis client and perform inference by reading the nodes from the model stored in Redis. A lambda invocation performs inference on the entire tree ensemble. Alternatively, multiple lambda invocations can perform inference concurrently on a subset of trees. The details and results are presented in \S ~\ref{lambda:conc}. 

%(MMTODO: describe the instance)

%be consistent in the formats
% only interleaved, classification, reg, class, reg
% Comment in the text about non-interleaved

\subsection{Embedded PACSET}\label{sec:pacsetemb}
%Rapid advances in technology have enabled the proliferation of 
Smart IoT devices are proliferating. It is estimated that by 2025, there will be more than 21 billion IoT devices\cite{iotfuture}. Devices often have limited computing and storage resources. An Arduino Uno consists of an ATmega328 microcontroller with a 6 MHz clock speed and only 32KB of flash memory and 4KB of SRAM. Thus, there are computational challenges to making these low resource devices ``intelligent'',  adding machine learning capabilities. 

Curent IoT deployments for machine learning require constant network connectivity. Models are deployed on cloud servers and IoT devices perform inference as a service through RESTful calls.
 This can be inefficient owing to large network latencies and is also prone to failure on networks with intermittent connectivity. Furthermore, there are security and privacy concerns. It is desirable to embed the models in the device. One line of research prunes machine learning models to fit them into device memory \cite{lowenergyrf}. Pruning results in performance loss. 
%a runtime-accuracy tradeoff because a pruned model is typically not as accurate as the original model.

Selective access removes the memory limitation. The entire model can be stored on external storage, such as a microSD card. Inference uses a diminimous amount of memory (a few megabytes) by accessing data one block at a time. 

\begin{table*}[ht]
 \begin{centering}
 \small
\begin{tabular}{llllll}
\hline
\textbf{Name} & \textbf{Type} & \textbf{Number of observations} & \textbf{Number of features} & \textbf{Number of Classes} & \textbf{Task} \\ \hline
CIFAR-10 & Image & 70000 & 32x32 = 1024 & 10 & RF Classification \\
Landsat & Image & 1000000 & 11 & 81 & RF Classification \\
Higgs  & Tabular & 1100000 & 28 & 2 & GBT Classification \\
Year  & Tabular & 515345 & 90 & N/A & RF Regression \\
WEC  & Tabular & 288000 & 49 & N/A & GBT Regression \\ \hline
\end{tabular}
\caption{Properties of the datasets used in  experiments. Datasets are publicly available in the UCI Machine Learning Library or the NASA website for Landsat.}
\label{tab:datasets}
\normalsize
\end{centering}
\end{table*}

\section{Experimental Evaluation}

%inidicate that io counting is always compared with real experiments. only proxy for how the algoirthm should work and we should always present real experiment.parnering between io counting experiments and system performance

We evaluate PACSET on a diverse set of machine-learning datasets, for both classification and regression, and for both gradient boosted trees and random forests. The primary metrics that we examine are inference latency for a single sample or small batches and memory footprint. We also look at batch inference throughput, to show that PACSET does not decrease throughput when used with large batches.  

We start with a general evaluation that examines performance with the internal memory in DRAM and external memory on SSD. This captures our deployment on larger than RAM models (\S \ref{sec:pacsetlarge}) for the datasets that have high dimensionality, a large number of classes, or many observations.  The evaluation compares analytical and measured results side by side. The measured results show system performance.  We compare this to {\em I/O counting} experiments that show the number of block transfers needed to run a workload in an external memory model. The comparison establishes that the system implementation tracks the analytical results closely and, thus, realizes the potential benefit of reducing I/O.  The experiments are run on AWS  EC2 c5d.large  instances  with 4GiB RAM, 2 vCPUs and 50 GB NVMe SSD for local storage. We ensure that all experiments work against a cold cache by using forests stored in different files in each iteration. This is much more computationally efficient than flushing the cache.
%We use the larger AWS r5dn.16xlarge instance for training large models.

We turn to specific deployment scenarios: 

\vspace{-10pt}

%\subsection{Infrastructure and Datasets}
%We describe the infrastructure used for each deployment scenario.
\begin{itemize}
\addtolength{\itemsep}{-5pt}
%    \item \textbf{PACSET on large memory models}: For inference, we used AWS EC2 c5d.large instances with 4GiB RAM and 2 vCPUs and 50 GB NVMe SSD local storage. We used larger instances for training large models specifically an r5dn.16xlarge instance.
 
    \item \textbf{PACSET as a service}: We perform inference on AWS Lambda functions with the default 128 MB RAM. Lamba functions access model data by making key/value read requests to an ElastiCache Redis Cluster implemented on 2 cache.m3.medium nodes.
% RB need more color here in the future.
 
    \item \textbf{Embedded PACSET}: We perform inference on a Raspberry Pi 2.   The Raspberry PI has a Broadcom BCM2836 900MHz quad-core ARM Cortex-A7 processor, 1 GB of SDRAM, a 64GB microSD card, and runs on 600mA at 5 V. 
    
%    \begin{table}[ht]
 %       \small
 %       \begin{tabular}{@{0.3\columnwidth}ll@{0.7\columnwidth}}
 %       \toprule
 %       \small
 %       \textbf{Model} & \textbf{Raspberry Pi 2} \\ \midrule
 %       CPU & Broadcom BCM2836 \\900MHz quad-core ARM \\Cortex-A7 processor \\
 %       RAM & 1 GB SDRAM \\
 %       USB Ports & 4 USB 2.0 ports \\
 %       Network & 10/100 Mbit/s Ethernet \\
 %       Power rating & 600 mA \\
 %       Power source & 5V Micro USB \\
 %       SD card & 4GB microSD card \\ \bottomrule
 %       \end{tabular}
 %       \caption{Raspberry Pi Hardware specifications}
 %       \label{tab:rpispecs}
 %       \normalsize
 %   \end{table}
 \end{itemize}
 
\vspace{-5pt}

%The description of the datasets used in our experiments are in  
Table~\ref{tab:datasets} summarizes the datasets on which we experiment and connects datasets to task (classification or regression) and data structure (random forest or gradient boosted tree).

%What question are you trying to answer?
%What experiment are you running to answer it?
%What are the expected results?
%What do the results show?
%If the results differ from your expectations, what did we learn?

% three sections
%General external memory experiment that mimic design section. Small datasets. CIFAR. 
%I/O block size in SSDs
%Specific experiments 
%Large models
%PACSET as a service
%embedded PACSET

%summary figure

%aggregate figure fashion mnist, cifar10, landsat
%Anchor figure, all the datasets, io count experiments, latency system, summary figure
%average latency
%merit of joy plot: see the distributions
%Early figure: violin plot and joy
%1: Joy plots for 1
%2: prefer violin
%3: joy plot
%5: Present top blue, present for 128 or 64., two panel figure
%table 1: bar chart: categories, BFS, DFS, WDFS, blockwise dfs, two bars
%table 2: error bars
\subsection{External Memory} %%MMTODO: Pick a title

\begin{figure}

     \vspace{-20pt}

     \includegraphics[width=\columnwidth]{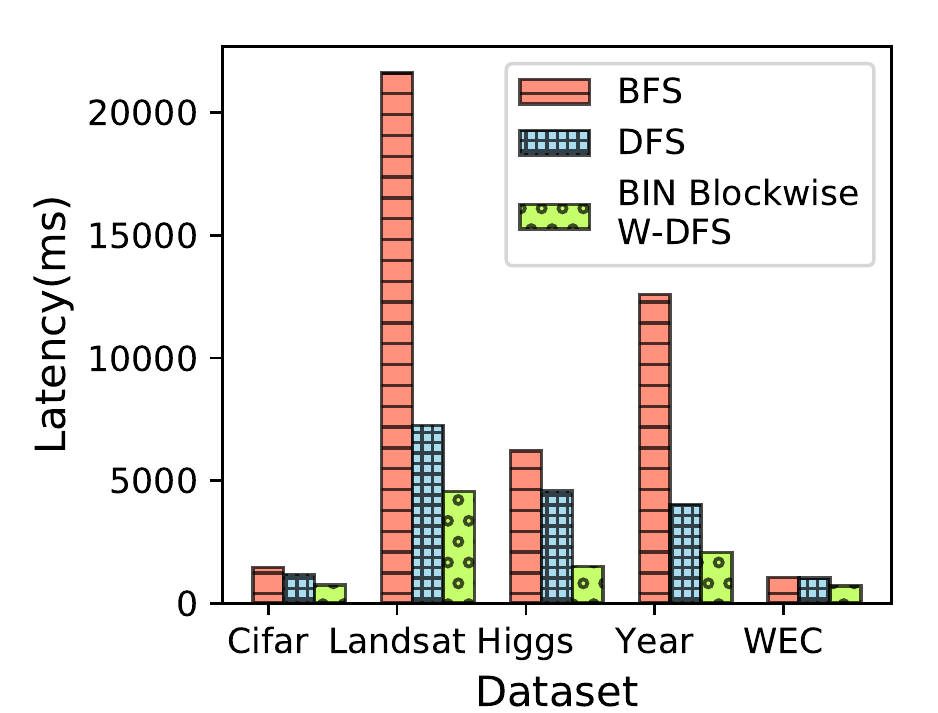}
     
          \vspace{-5pt} 
          
     \caption{Latency to perform a single inference in an tree emsemble that compares PACSET with all optimizations (BIN blockwise WDFS) with the BFS and DFS layouts.}
     \label{fig:aggfig}
     
     \vspace{-10pt} 
\end{figure}

Our first experiment captures overall performance for big datasets.  It measures the inference latency of PACSET with all optimizations against the baseline BFS layout used by XGBoost \cite{chen2016xgboost} and DFS layout used by scikit-learn \cite{pedregosa2011scikit}.  For all layouts, we use a JSON format for model representation.\footnote{We aspired to use more optimized serialization formats, such as {\sf capnproto}, but found that they were slower than JSON and had incompatibilities with ARM processors.}  For the larger models, Landsat, Higgs, Year, PACSET sees a 2-6 times reduction in latency (Figure~\ref{fig:aggfig}). This experiment stripes 682 tree across the BIN. This number is chosen so that the top two levels of the tree fit in a single block. The 64K block size is determined by the minimum I/O size of 4 KB multiplied by the 16 parallel channels of the device, resulting in 2048 tree nodes per block.

For smaller models, there is less benefit owing to the larger block size of enterprise SSDs. Small models fit into few blocks and selective access is less effective.  In the limit, an ensemble that fits into one block, PACSET would provide no savings.  Smaller models  see more savings with finer-grained I/O, i.e.~in the cloud and on embedded devices. 

We next compare against the implementation of inference in scikit-learn and XGBoost, comparing performance at different batch sizes and the memory footprint (Table~\ref{tab:baselinecomparison}). scikit-learn loads the entire models into memory before performing inference. This process is slow because models have a large number of trees. Loading the entire model is best for large batches that search many paths and touch all blocks in any layout.  Loading the model (I/O only) takes 15s. Scikit-learn performs additional computation that scales with the batch size. I/O dominates performance.  This experiment used the CIFAR-10 dataset with a random forest of 682 trees. This dataset occupies 3.5 GB of memory. Larger datasets overflow memory in scikit-learn.

The selective access of PACSET accesses only the data that are needed, resulting in much lower latency for small batches and reduces the memory footprint by orders of magnitude. Memory usage is calculated using {\sf htop} at 500 ms intervals. {\sf htop}
does not account for anonymous or pinned pages.  The 0.003 GB of memory is the allocated memory and PACSET work correctly in that amount of memory. PACSET performance may benefit slightly from the reuse of anonymous pages, but this only happens within each batch because we ensure a cold cache. Although the overall performance seems slow, 6  seconds for a batch size of 10 samples, inference accesses more than 80,000 nodes from SSD, using less than 1 ms per block I/O. 

Inference on large batches of 2000 show that PACSET performs much worse that loading the entire model.  This is expected. Large batches access all blocks in the model. PACSET pages these in on demand. This will always be less efficient than performing sequential I/O.

%In this section we present the main results in a summary figure. We would like to compare the effect of the packing layouts described in \S~\ref{packingalgo} on the I/O count and consequently inference latency. The I/O count is simply an estimate of the number of I/O blocks that need to be read to perform inference, keeping in mind that multiples of 4KB are read from the disk at a time. The inference latency is the time taken to perform inference on one test observation sample. The inference latencies should be directly proportional to the I/O counts. This is because on average, the time taken for random access of any block in the file should be around the same. Thus, the total inference time should be approximately equal to the I/O count times the average block access latency. The algorithm is a random forest trained on 682 trees. The inference is performed on an EC2 instance and the model is stored on an SSD with a page size of 64KB. This corresponds to 2048 tree nodes (each node is 32 bytes long). We choose the number of trees so that the first two levels(including the root) of all the trees in the ensemble fit in a block. In Figure~\ref{fig:aggfig}, we present bar graphs to visualize the same. We compare the best layout (interleaved BIN + blockwise) with the baseline layout(BFS). We split the figure into different sub bargraphs such that there is one bar graph per dataset. Each bar graph consists of two bars, one showing the average I/O counts and the other for the average latencies. We see that the I/O counts roughly correspond to the latencies. 

\begin{table*}[ht]\
\begin{center}
\small
\begin{tabular}{@{}lllll@{}}
\toprule
\textbf{Package} & \textbf{Time to load model (s)} & \textbf{\begin{tabular}[c]{@{}l@{}} Inference Latency (s)\\ Batch size = 10\end{tabular}} & \textbf{\begin{tabular}[c]{@{}l@{}} Inference Latency (s)\\ Batch size = 2000\end{tabular}} & \textbf{ Memory (GB)}\\ \midrule

scikit-learn & 14.98 & 17.74 & 18.052 & 3.5 \\
PACSET & N/A & 1.041 & 61.449 & 0.003 \\ \bottomrule
\end{tabular}
\caption{Table comparing PACSET with scikit-learn on inference latency in addition to memory used}
\label{tab:baselinecomparison}
\end{center}
\normalsize
\end{table*}

         \def\nameslatency{
        {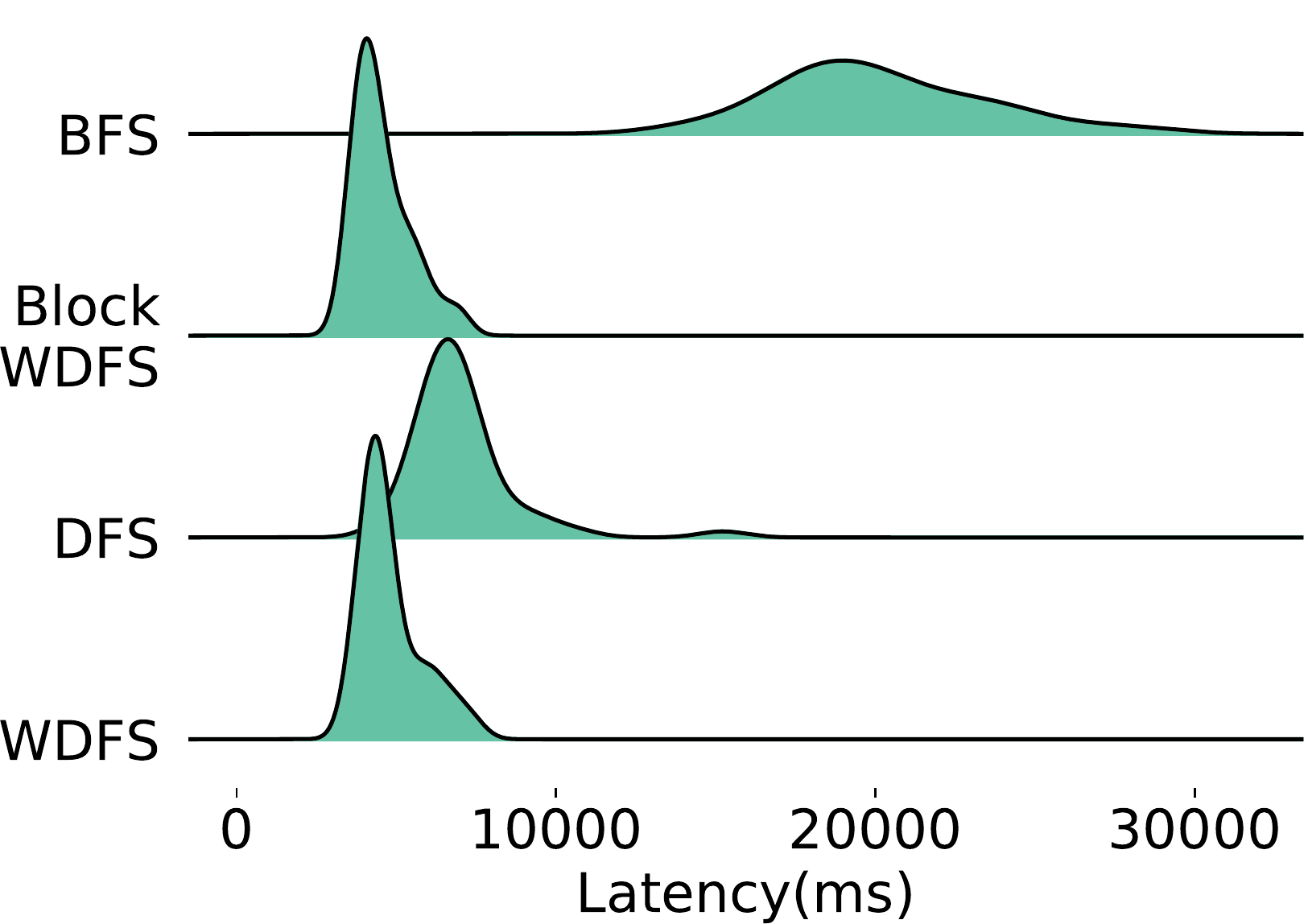},
        {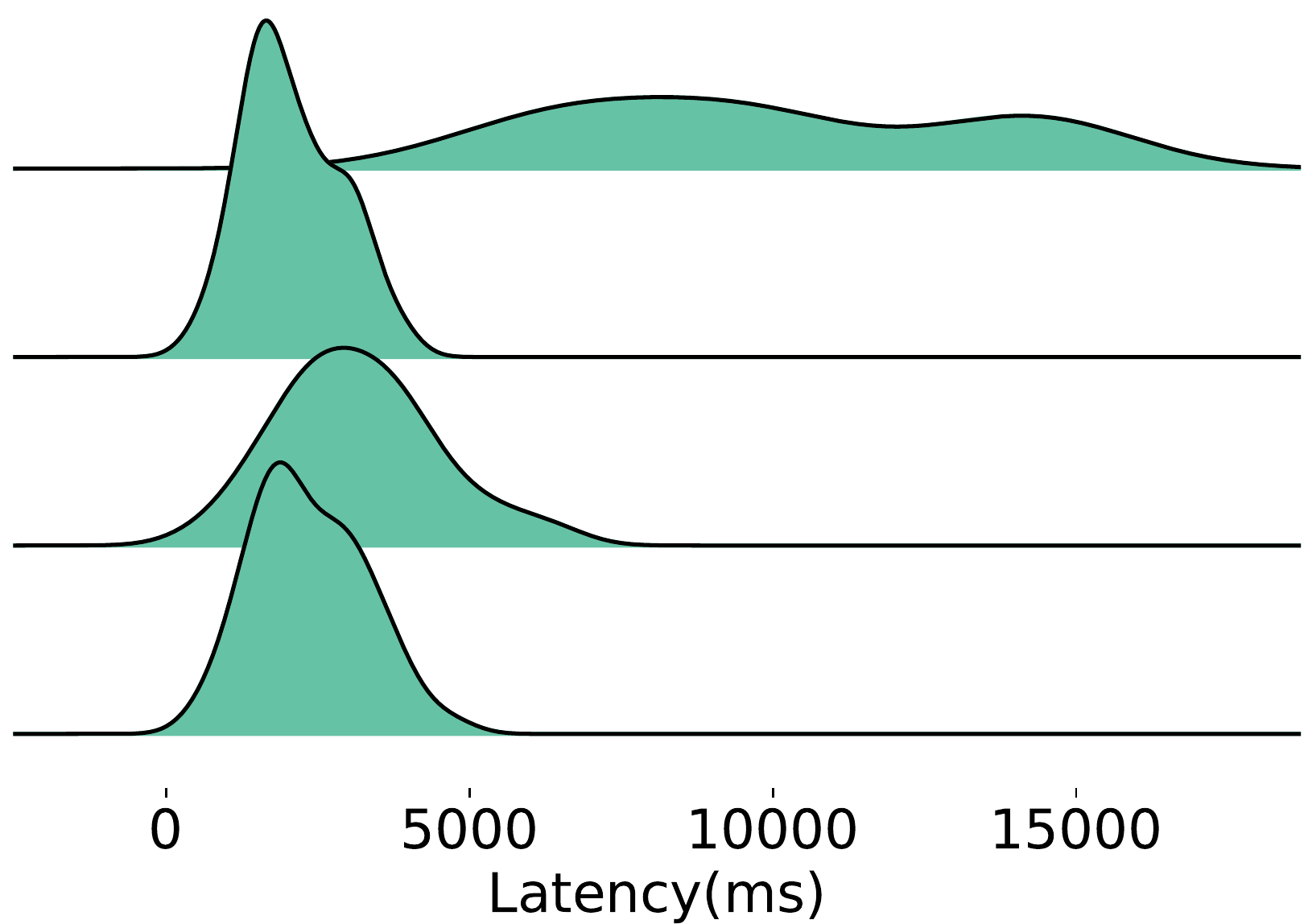},
        {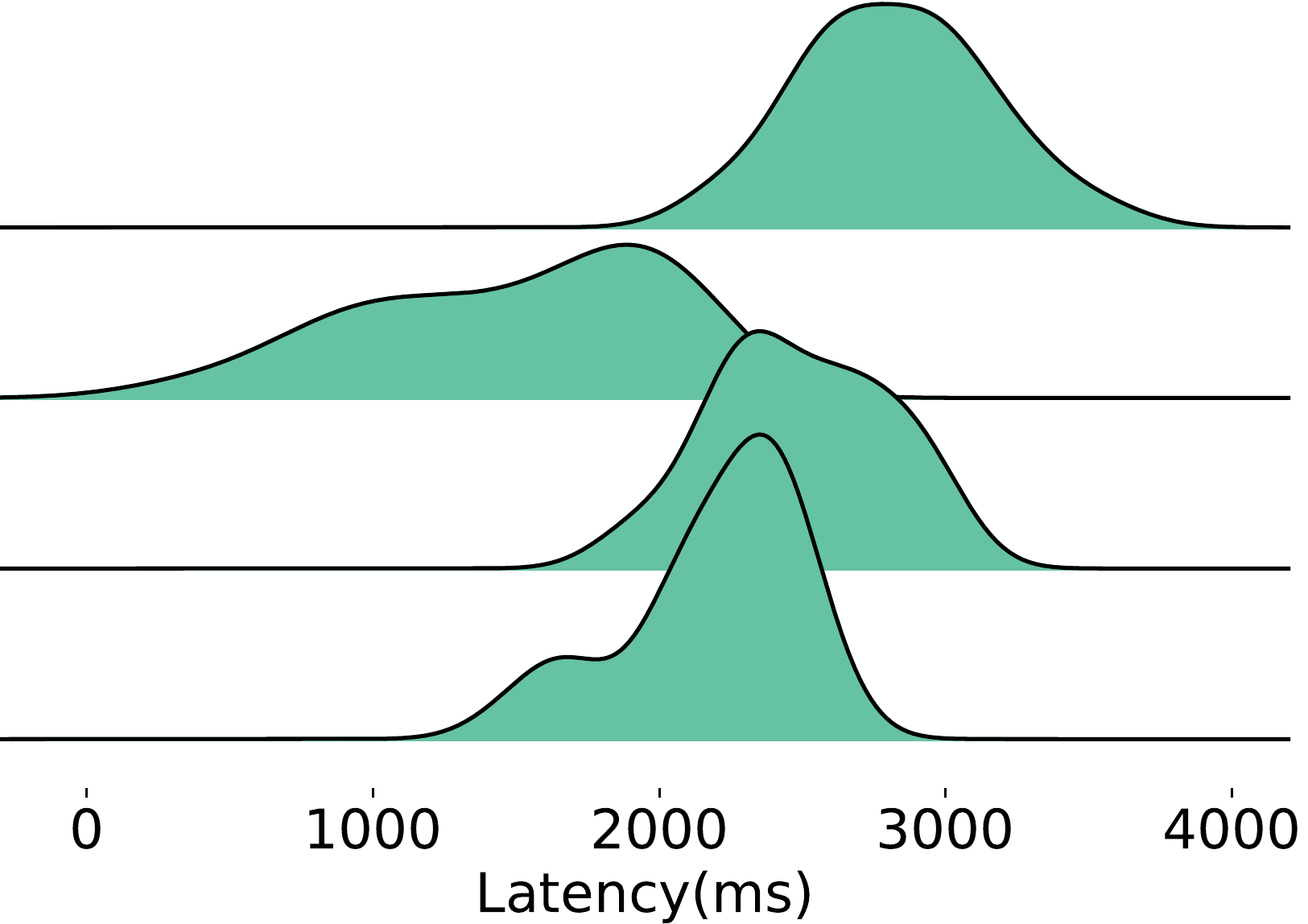},
        {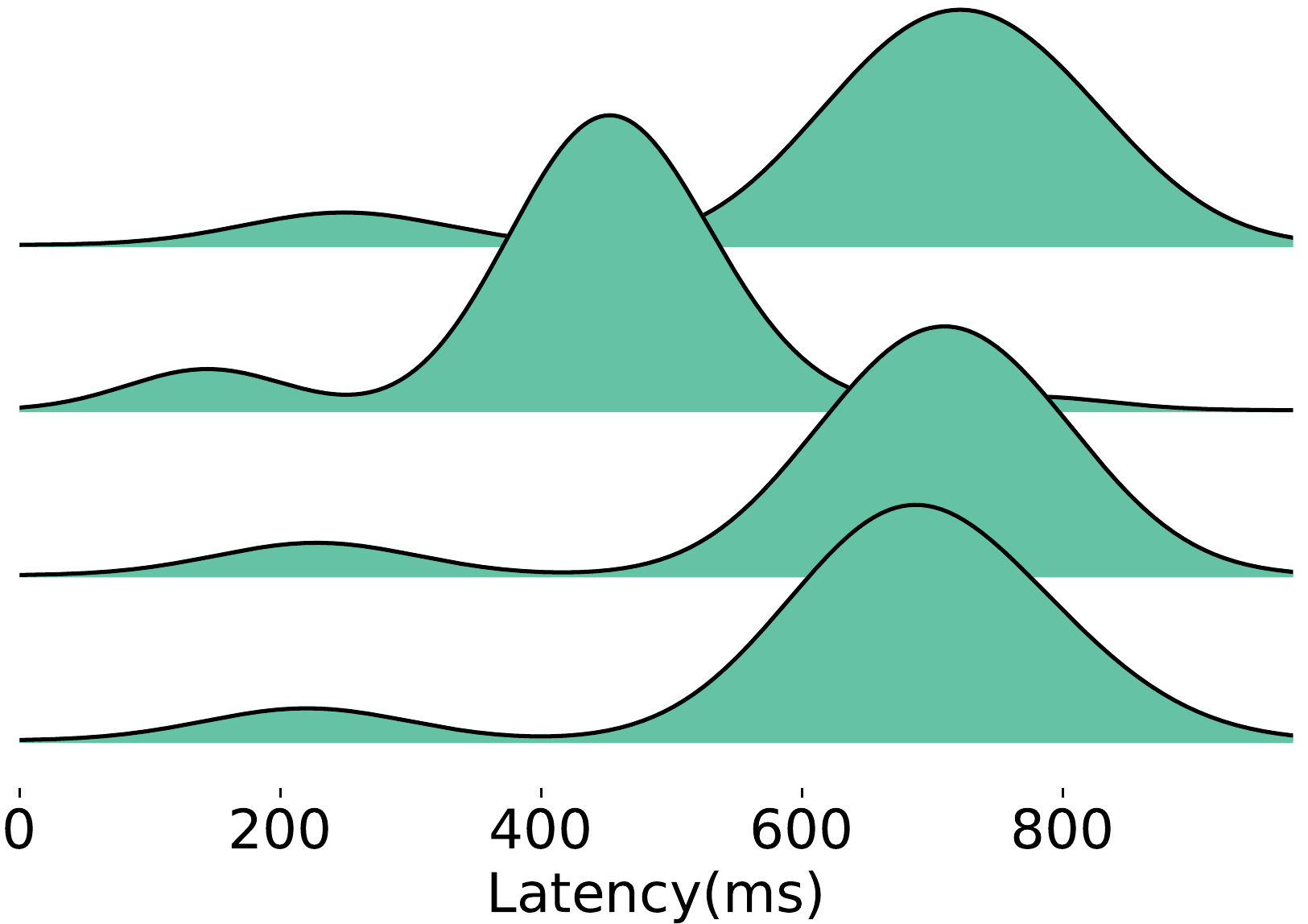}}
    
    \begin{figure*}
        \foreach \name in \nameslatency {%
             \begin{subfigure}[p]{0.47\columnwidth}
                 \includegraphics[width=1.07\textwidth]{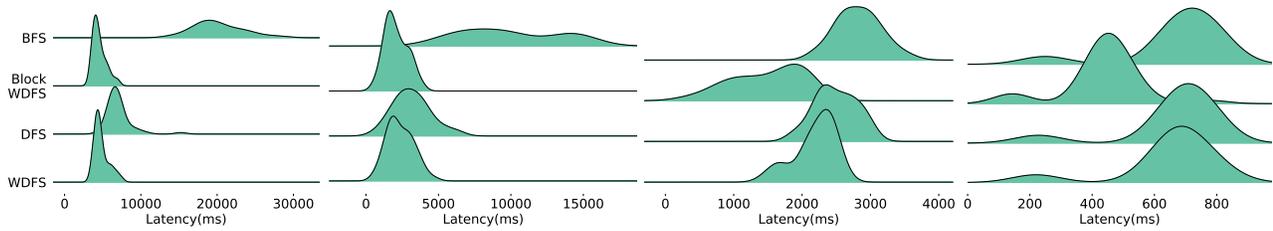}
                 
             \end{subfigure}\quad
        }
        \caption{Inference latency as a function of forest layout for RF classification (left), RF regression (left-center) , GBT classification (right-center) and GBT regression (right). }\label{fig:layoutlat}
    \end{figure*}   
    \def\names{
        {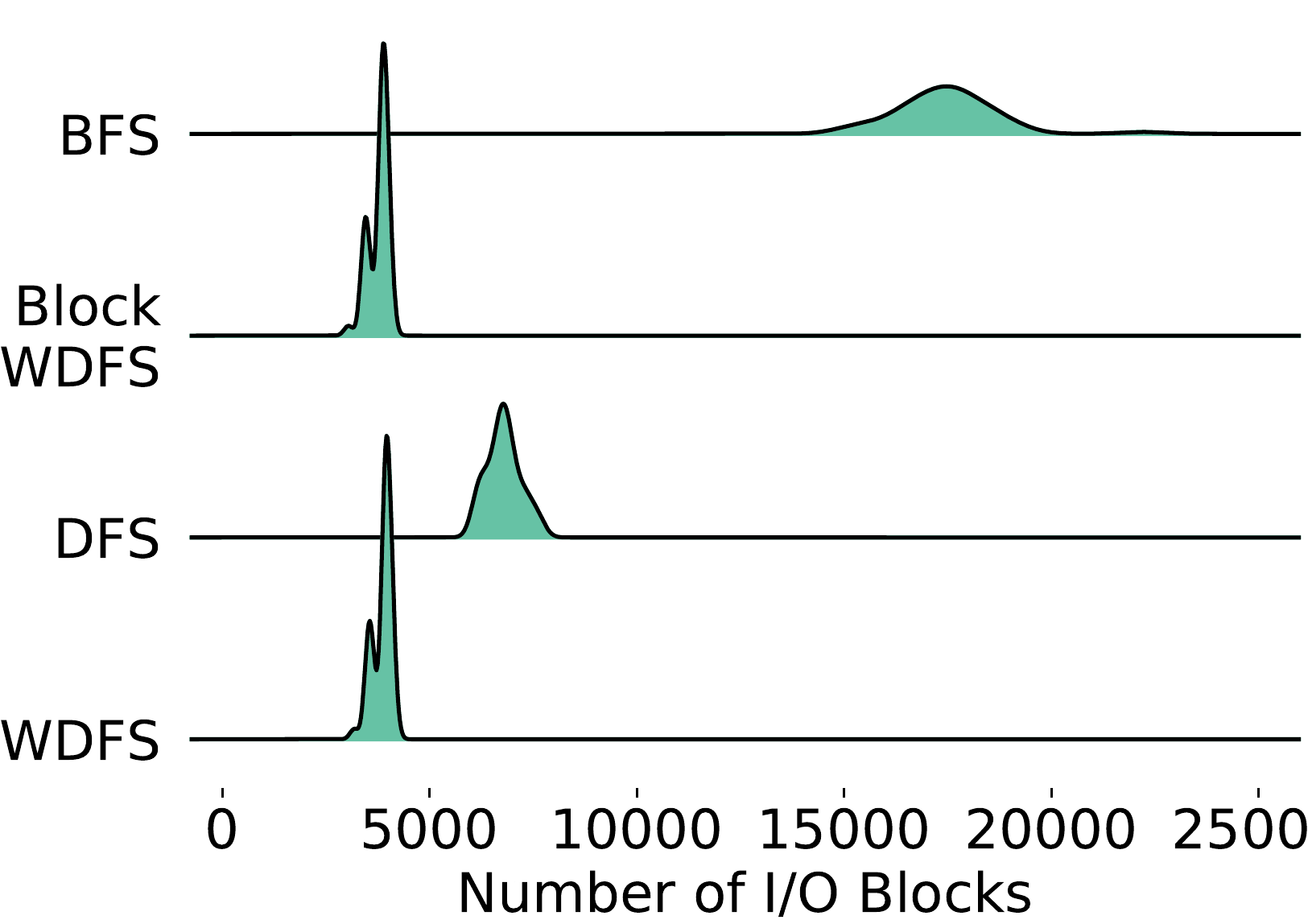},
        {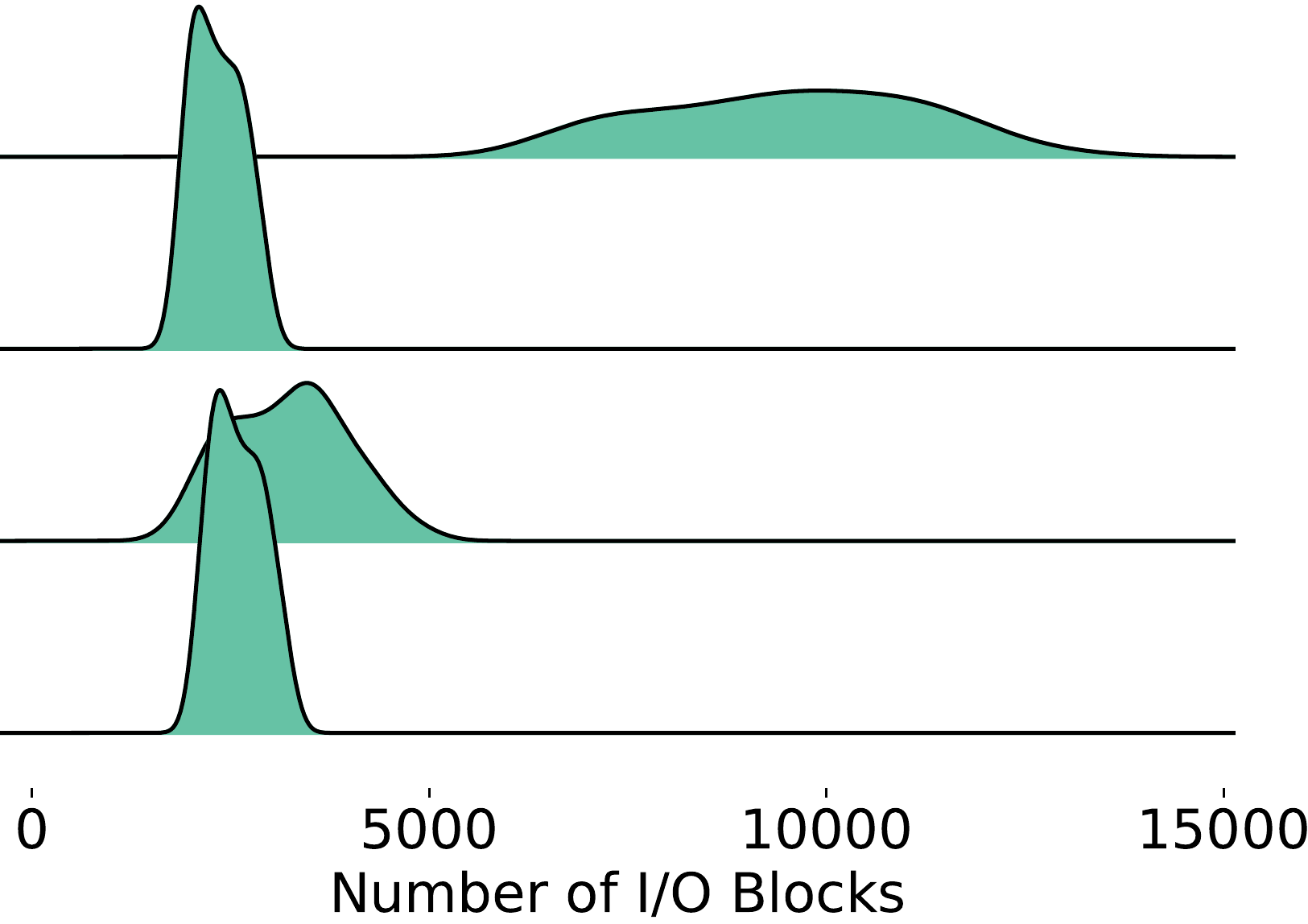},
        {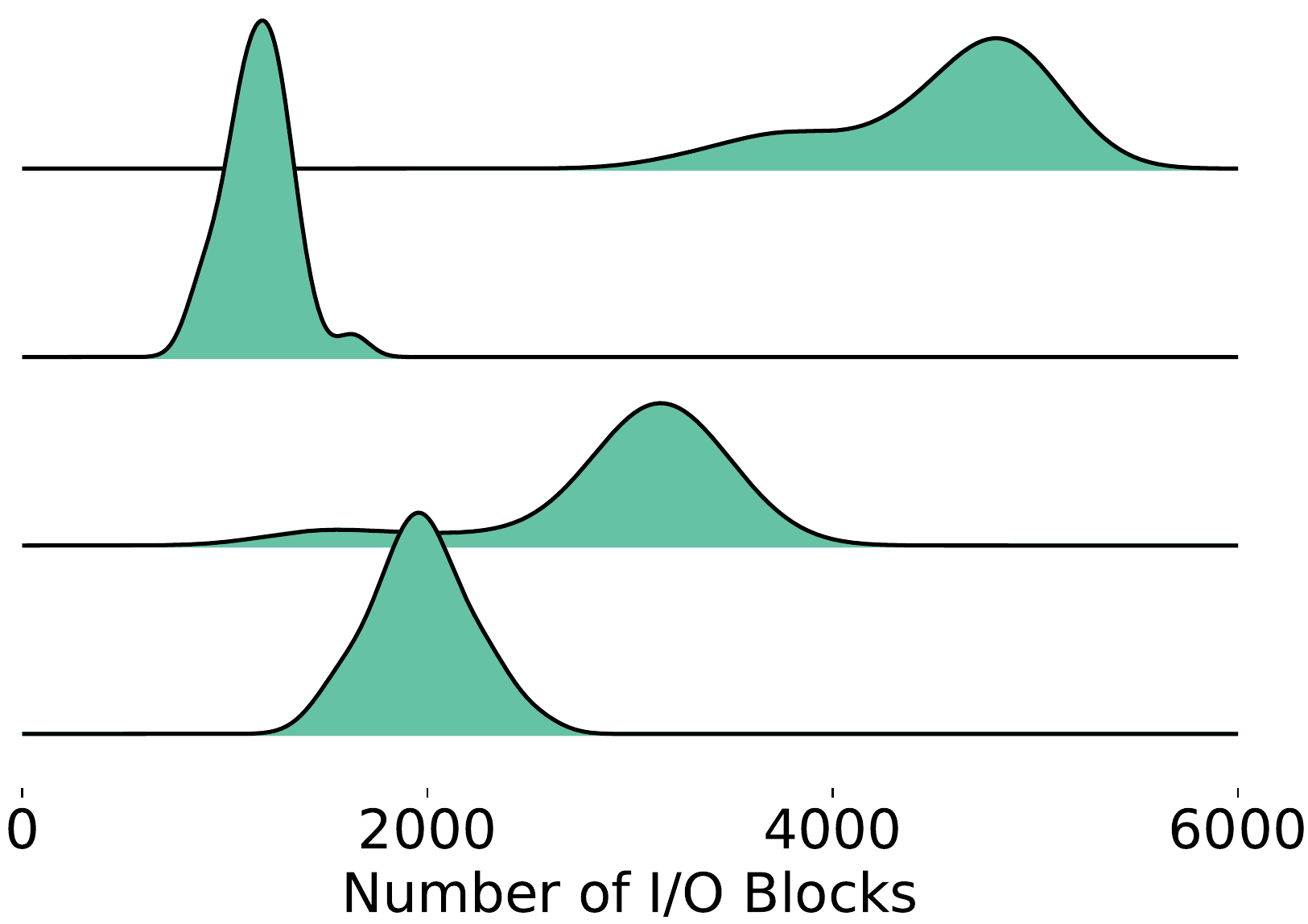},
        {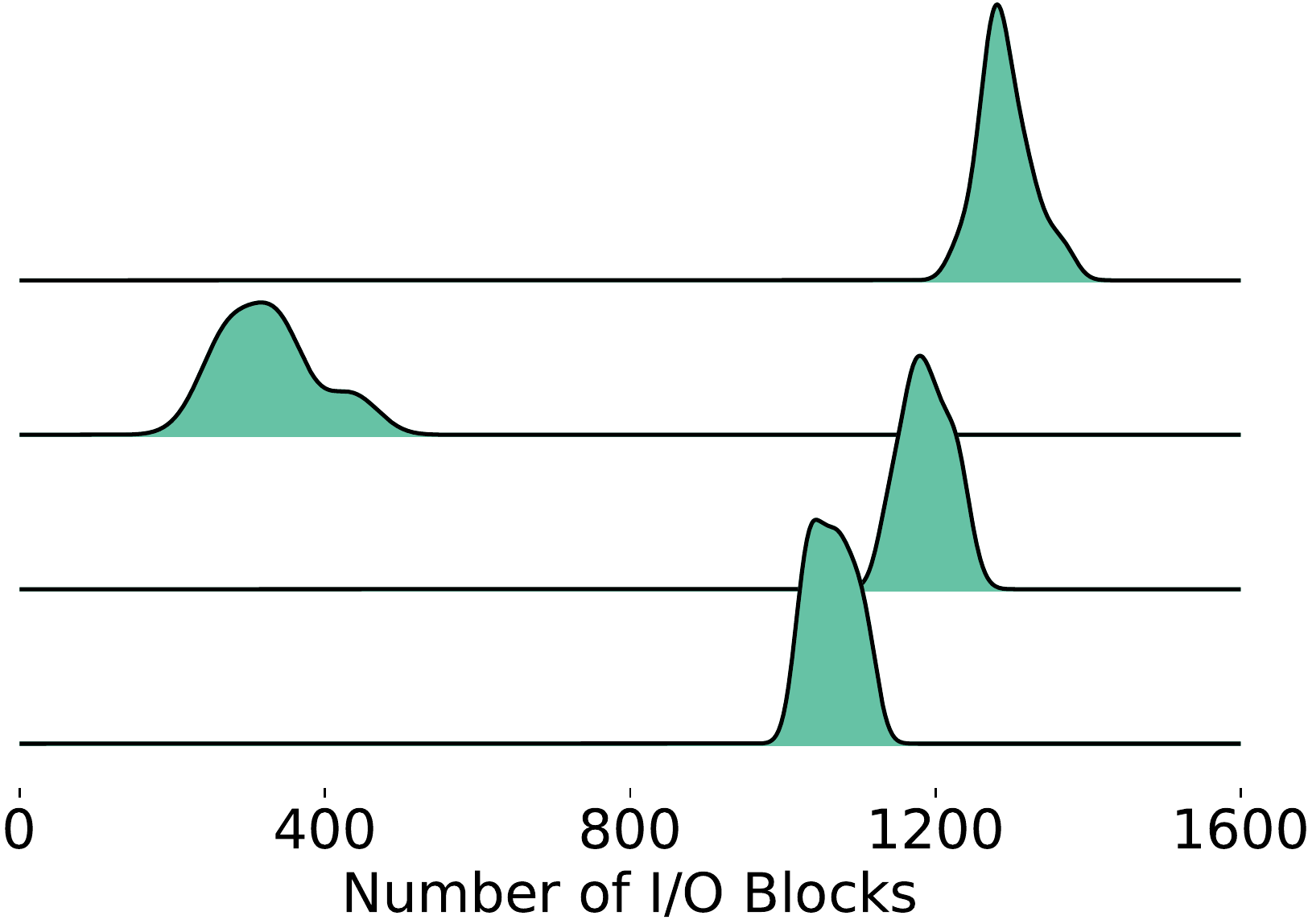}}
    
    \begin{figure*}[h!]
        \foreach \name in \names {%
             \begin{subfigure}[p]{0.47\columnwidth}
                 \includegraphics[width=1.07\textwidth]{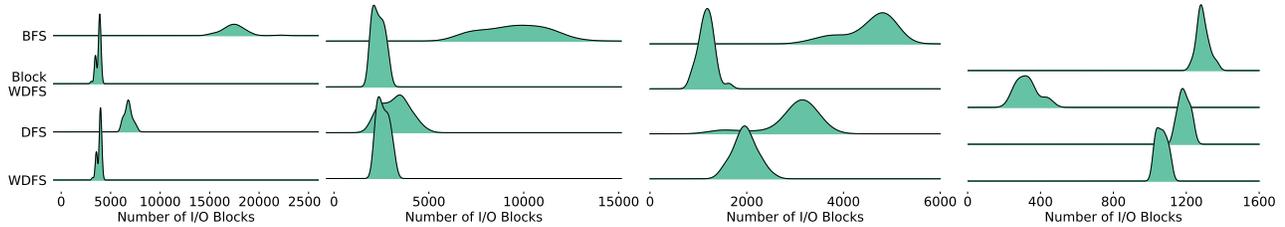}
             \end{subfigure}\quad
        }
        \caption{Distribution of the I\slash O block counts as a function of forest layout for RF classification (left), RF regression (left-center) , GBT classification (right-center) and GBT regression (right).}\label{fig:blockcount}
    \end{figure*}

\noindent {\bf Classification and Regression}

%\subsection{PACSET I/O Counting Experiments} %\label{exp:iocount}
%MMTODO: maybe rename ?

PACSET post-processes trained ensembles to improve inference latency for different data structures and learning goals. We measure performance for  regression and classification in random forests (RF) and gradient boosted trees (GBT).  

PACSET reduces inference latency in all combinations (Figure~\ref{fig:layoutlat}.  These experiments demonstrate the the techniques apply across a range of learning tasks. All layouts use interleaved bins. Thus, they do not have the 2x-6x performance improvement seen over unoptimized BFS and DFS. 

For random forests, the cardinality weighting of WDFS provides almost all of the benefit. Random forests train an ensemble of 682 trees and do not limit tree depth. Block WDFS performs best in all cases and is more than a 50\% improvement over unweighted DFS.

For gradient-boosted trees, the smaller tree size makes block-alignment more important. Gradient boosted trees train an ensemble of 2048 trees with a max depth of 12.
Gradient boosted trees typically employ a higher number of ``short'' trees when compared with random forests \cite{chen2016xgboost}. The residuals are smaller than a block and aligning residuals to block boundaries avoids many I/Os. 

We compare measured performance against analytical results (Figure \ref{fig:blockcount}). These experiments run the same inference workload and count the number of unique blocks accessed. This represents the minimum number of block transfers from an external memory to run a workload. The number of I/Os represents a lower bound on latency; it assumes that once a block is accessed it is always available in cache.

System latency experiments track I/O counting analytics closely, showing that PACSET realizes much of the performance benefits of I/O reduction. The system measurements have wider distributions,  owing to caching and scheduling/skew. The random forest results track I/O counting closely.  The gradient boosted forest results show that do not track as closely. The smaller trees result in discretization and alignment effects that reduce the cache hit rate.

        \begin{figure}[ht]
          
            \includegraphics[width=\columnwidth]{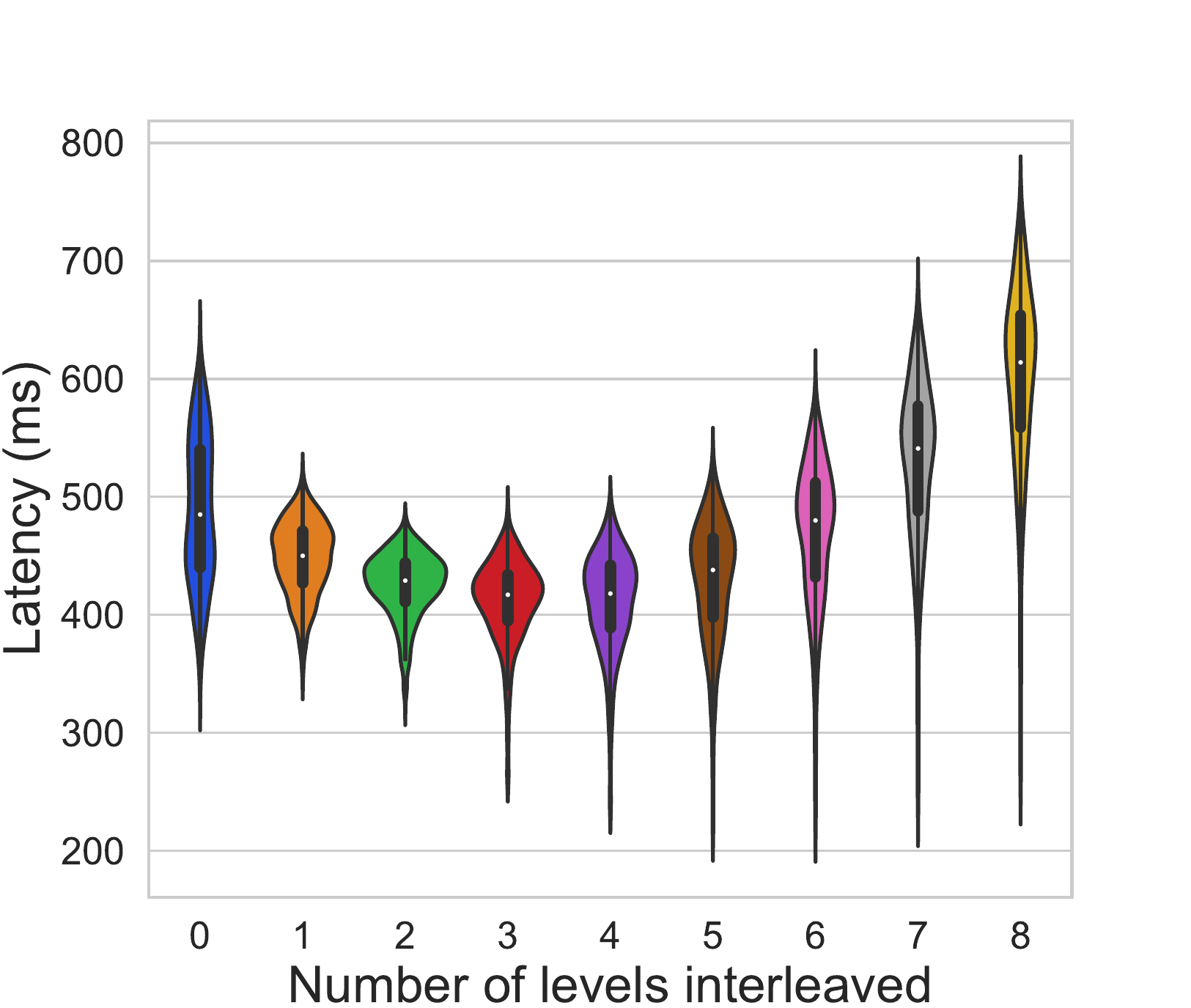}
            \caption{Distribution of latency as a function of the depth of the interleaved bins in PACSET.}\label{fig:bindepth}
        \end{figure}   
    
{\noindent \bf Depth of interleaved bins}

Beacuse much of the I/O savings comes from the static striping of high-level nodes in the tree, we explore the bin depth hyperparameter.  All prior experiments ran with a bin depth of two. This section justifies that decision. It also shows that shallow bins reduce I/O the most and that the WDFS layout is needed to handle deep trees.
Figure \ref{fig:bindepth} shows inference latency as a function of the bin depth for a random forest of 128 trees on the CIFAR-10 data set. This experiment varied the bin width---number of trees in a bin---proportionally, so that the bin fits within an I/O block. Here a bin depth of 3 has the lower mean and a bin depth of 2 has the smallest variance. For the high levels in the tree, striping increases the density of access.  Every root node is used, every other level 1 node, etc. For datasets that are more evenly distributed across leaves, such as CIFAR, deeper bins are preferable.  For datasets that are more skewed, such as Landsat, shallower bins are preferable because popular paths through the tree result in higher density of nodes per I/O. Overall, the best choice is two or three and we choose two because variance is always smaller, it is often lower latency, and because two-levels bins of 682 fit almost exactly within a single block.

       \begin{figure}[t]
                \includegraphics[width=0.5\textwidth]{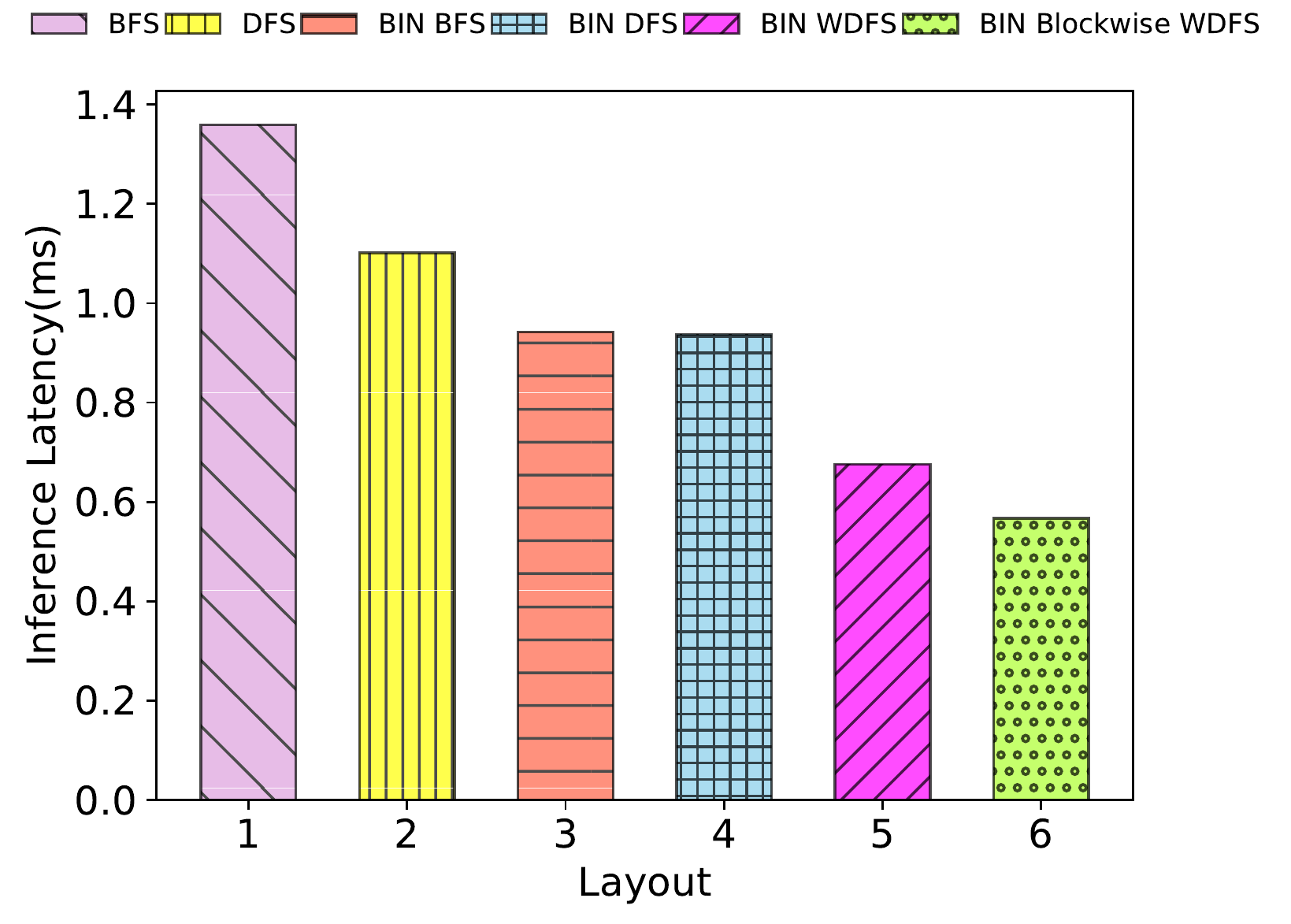}
                
                    \vspace{-8pt}
                    
                \caption{Cold start inference latency as a function of layout for the CIFAR-10 dataset.}
                \label{fig:lambdalayout}
         \end{figure}

\subsection{PACSET as a service}\label{exp:pacsetaas}
 
We deploy tree-ensemble inference as a cloud microservice on AWS Lambda and show that we can realize subsecond inference latency.  We store the models in the Redis key value store, using a unique block identifier as the key and an array of node data as the value.  Each key/value pair corresponds to a block of nodes in external memory. Figure~\ref{fig:lambdalayout} demonstrates that PACSET with all optimizations provides a 2.5 times reduction in latency when compared with BFS and over 2 times when compared with DFS.  Most of the reduction can be attributed to interleaved bins. Examining the different layouts reveals that block alignment and weighted DFS provides reduces latency by about factor of 40\% over binned DFS and BFS.  

For all datasets and learning tasks, PACSET performs cold-start inference in less than a second, which makes it viable to deploy as a classification or regression service in a Web application.
Latency measurements include the Lambda invocation overhead, i.e.~the time to allocate and start the compute resource, which is called a ``cold start''. PACSET performs inference for CIFAR-10 in less than 600ms. All layouts pay the same startup cost for cold start, which we measured at about 100 ms. This overhead reduces the possible performance improvement for PACSET in the cloud when compared with PACSET on SSDs.

Additional experiments show that optimizations apply over a wide range of parameters.
In-memory key/value stores provide fine-grained I/O and low-latency access. An empirical evaluation of hyperparameters determine that a bin depth of two and a block size of only 8 nodes per key/value pair minimizes latency. This is in contrast to a block size of 2048 for external memory on enterprise SSDs. We report on these experiment in the Apppendix. The appendix also reveals that cloud-microservices are I/O bound and do not benefit from parallelizing inference at a fine granularity. The best policy deploys a Lambda function per bin.

\subsection{Embedded PACSET}

\begin{figure}
    \includegraphics[width=0.9\columnwidth]{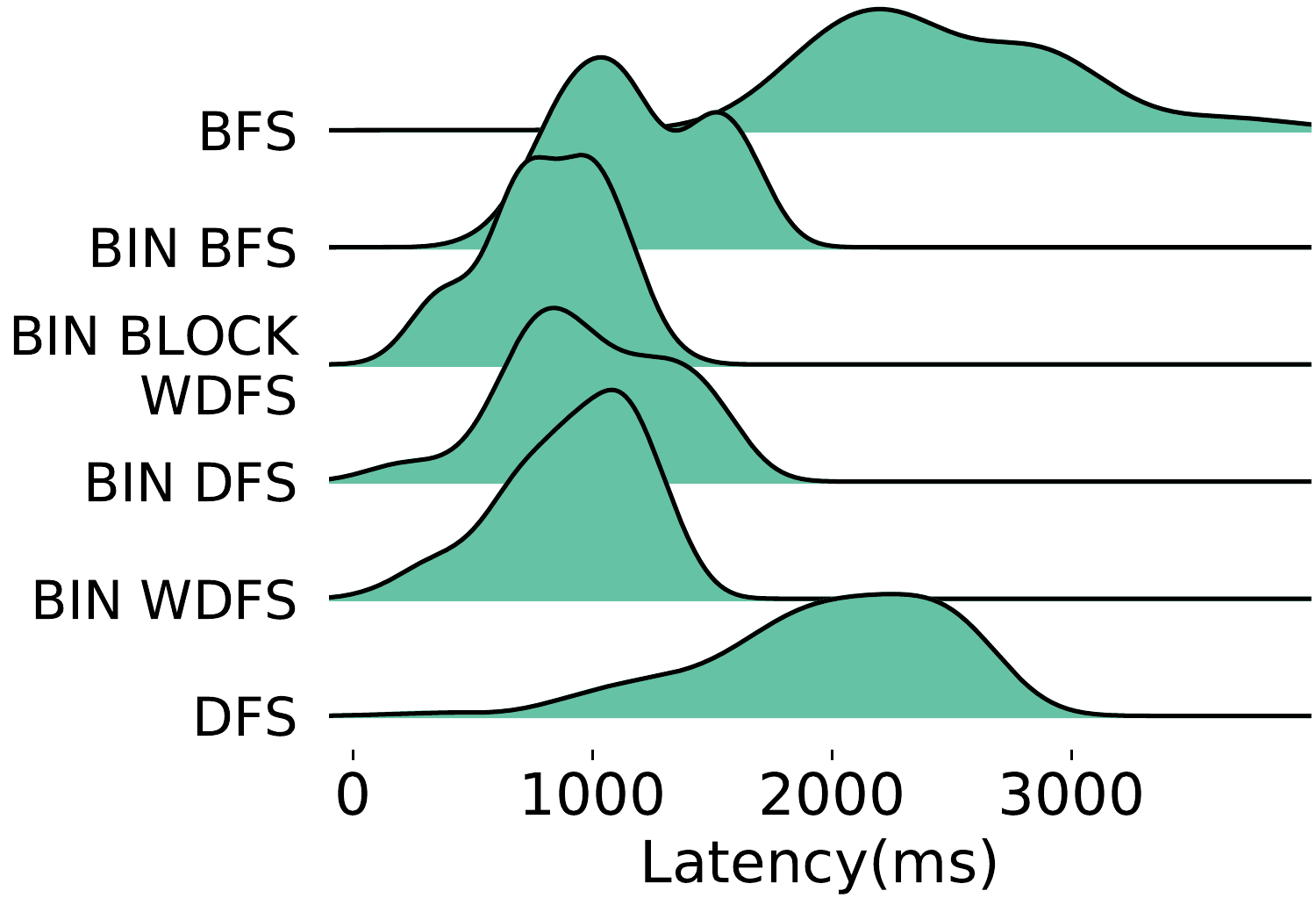}
    \caption{Layout versus latencies for the CIFAR-10 dataset on embedded PACSET with an interleaved BIN depth of 2.}
    \label{fig:pacsetemb}
    \vspace{-10pt}
\end{figure}

The PACSET layout cuts the inference latency on embedded devices by a factor 2.5 when compared with BFS and DFS on the CIFAR-10 data set (Figure \ref{fig:pacsetemb}). 
This experiment trained a random forest of 128 trees on the Raspberry PI and packed the forest with a bin depth of 2. This is a favorable execution environment that shows better latency reduction than  SSDs or microservices, because there are no startup costs and there is larger I/O latency gap between RAM and microSD storage.  The benefit on CIFAR-10 on SSDS was only 2.0 times. The I/O gap magnifies the benefit of reducing I/O.

%and is well larger than the 1 GB of internal memory. 

The small 4 kb block size of the microSD card makes block alignment critical to realize performance. 
The distribution of inference latencies on CIFAR-10 reveal that WDFS alone provides little benefit over binned BFS and DFS, whereas block aligned PACSET reduces latency by 20\%.

%In this section we describe the experimental setup and results for PACSET deployed on an embedded device. We conduct all our benchmark on the Raspberry Pi model 2. The hardware specs are listed in Table\ref{tab:rpispecs}. We store the model on a microSD card with 64GB of memory. However, our methods are general and can be deployed on any embedded device with limited on-device RAM and support for inserting microSD card large enough to fit the model. The question we want to answer here is the following. Do our packing layouts yield lower inference latencies when the model is stored in a microSD card and inference occurs on the embedded device? To answer this question we repeat the layout experiment described in \S\ref{exp:layout} but on the embedded device. We train a random forest model with 128 trees on the CIFAR-10 dataset. 

%We see in Figure \ref{sec:pacsetemb} that our blockwise packing layout shows a greater reduction in inference latency in this setting than in the previous cases. This is because the latency of reads from a microSD card are much lower than an SSD. I/O savings are more distinct as we see a greater discord between external and main memory read latencies.

%\printAffiliationsAndNotice{}  % leave blank if no need to mention equal contribution
%\printAffiliationsAndNotice{\mlsysEqualContribution} % otherwise use the standard text.

\section{Conclusions}

Latency is the key performance metric in interactive machine learning that performs inference on demand.  Our research drives latency down for classification and regression with gradient boosted trees and random forests. These methods are well suited to distributed deployments because tree ensembles requires little computation and and a small amount of memory.

With latency reduction as the goal, PACSET creates serialization formats that support selective I/O---accessing only the portions of the forest needed to perform inference on a single observation or a small batch.  This is a straightforward goal, but it diverges from existing systems that are optimized for large batches and load the entire model into memory.  Moving to selective I/O provides order of magnitude latency benefits, but this is in comparison to systems that were designed for batch workloads. 

PACSET refines the performance of selective access using principles from external memory algorithms that minimize the number of block transfers from an external memory to RAM.  We optimize the placement of trees nodes in the forest into memory to encode access locality, identifying static relationships across mutliple trees and dynamic/statisical relationships with trees. This process reduces latency by a factor of 2 to 6 compared with standard external memory layouts that serialize one tree at a time.

This work opens several research questions. Our efforts share goals with techniques that prune or compress models. Combining model reduction techniques with packing would be beneficial and is non-trivial. Also, our efforts focus on tree ensembles, but the fundamental principle of identifying and encoding data relationships could apply to any ensemble. This leads to the question: Can we implement a generic machine learning model storage framework for latency reduction? Doing so for deep neural networks would be am important first step toward generality.

%into   particularly those that intergrate classification

%In this paper we proposed a layout and system design for efficient inference from a tree ensemble when the model is stored in external memory. We described three deployment use cases and presented experimental results to corroborate our findings. We showed that in these scenarios, instead of reading the entire ensemble into memory, it can be helpful to perform selective access. Selective I/O access entails reading parts of the model from external memory as and when needed. We then described a layout, utilizing statistical properties of node cardinality and knowledge of the blocksize to minimize the number of I/Os required to to perform inference.  Our layouts result reduction in latency by over an order of magnitude and up to 6 times over a naive breadth first layout of trees. One line of further investigation is to see if we can incorporating model compression into PACSET. Another deployment use case worth investigating is when the model doesn't fit into GPU memory. Finally, it would be interesting to see if layout techniques work and selective access can be applied to general ensemble models and not just tree ensembles.

\bibliography{pacset.bib}

\begin{thebibliography}{34}
\providecommand{\natexlab}[1]{#1}
\providecommand{\url}[1]{\texttt{#1}}
\expandafter\ifx\csname urlstyle\endcsname\relax
  \providecommand{\doi}[1]{doi: #1}\else
  \providecommand{\doi}{doi: \begingroup \urlstyle{rm}\Url}\fi

\bibitem[Alafate \& Freund(2019)Alafate and Freund]{alafate2019faster}
Alafate, J. and Freund, Y.~S.
\newblock Faster boosting with smaller memory.
\newblock In \emph{Advances in Neural Information Processing Systems}, pp.\
  11367--11376, 2019.

\bibitem[Ananthanarayanan et~al.(2017)Ananthanarayanan, Bahl, Bod{\'\i}k,
  Chintalapudi, Philipose, Ravindranath, and Sinha]{ananthanarayanan2017real}
Ananthanarayanan, G., Bahl, P., Bod{\'\i}k, P., Chintalapudi, K., Philipose,
  M., Ravindranath, L., and Sinha, S.
\newblock Real-time video analytics: The killer app for edge computing.
\newblock \emph{Computer}, 50\penalty0 (10):\penalty0 58--67, 2017.

\bibitem[Anghel et~al.(2019)Anghel, Ioannou, Parnell, Papandreou,
  Mendler-D{\"u}nner, and Pozidis]{anghel2019breadth}
Anghel, A., Ioannou, N., Parnell, T., Papandreou, N., Mendler-D{\"u}nner, C.,
  and Pozidis, H.
\newblock Breadth-first, depth-next training of random forests.
\newblock \emph{arXiv preprint arXiv:1910.06853}, 2019.

\bibitem[Anonymous()]{anonymous}
Anonymous.
\newblock Anonymous.

\bibitem[Breddels(2014)]{vaex:2009}
Breddels, M.~A.
\newblock \emph{Vaex}, 2014.
\newblock URL \url{https://docs.vaex.io/en/latest/tutorial.html}.

\bibitem[Cai et~al.(2017)Cai, Ren, Zhang, Malialis, Wang, Yu, and
  Guo]{cai2017real}
Cai, H., Ren, K., Zhang, W., Malialis, K., Wang, J., Yu, Y., and Guo, D.
\newblock Real-time bidding by reinforcement learning in display advertising.
\newblock In \emph{Web Search and Data Mining}, pp.\  661--670, 2017.

\bibitem[Carreira et~al.(2019)Carreira, Fonseca, Tumanov, Zhang, and
  Katz]{10.1145/3357223.3362711}
Carreira, J., Fonseca, P., Tumanov, A., Zhang, A., and Katz, R.
\newblock Cirrus: A serverless framework for end-to-end {ML} workflows.
\newblock In \emph{ACM Symposium on Cloud Computing}, pp.\  13–24, 2019.
\newblock \doi{10.1145/3357223.3362711}.

\bibitem[Chen \& Guestrin(2016)Chen and Guestrin]{chen2016xgboost}
Chen, T. and Guestrin, C.
\newblock {XGBoost}: A scalable tree boosting system.
\newblock In \emph{ACM SIGKDD International Conference on Knowledge Discovery
  and Data Mining}, pp.\  785--794, 2016.

\bibitem[Cho \& Li(2018)Cho and Li]{cho2018treelite}
Cho, H. and Li, M.
\newblock Treelite: toolbox for decision tree deployment.
\newblock 2018.

\bibitem[Crankshaw et~al.(2017)Crankshaw, Wang, Zhou, Franklin, Gonzalez, and
  Stoica]{crankshaw2017clipper}
Crankshaw, D., Wang, X., Zhou, G., Franklin, M.~J., Gonzalez, J.~E., and
  Stoica, I.
\newblock Clipper: A low-latency online prediction serving system.
\newblock In \emph{{USENIX} Symposium on Networked Systems Design and
  Implementation}, pp.\  613--627, 2017.

\bibitem[Eads et~al.(2018)Eads, Baines, and Bloom]{eadsmemory}
Eads, D., Baines, P., and Bloom, J.~S.
\newblock Memory-efficient data structures for learning and prediction.
\newblock In \emph{Machine Learning and Systems}, 2018.

\bibitem[Feng et~al.(2018)Feng, Yu, and Zhou]{feng18mlgbt}
Feng, J., Yu, Y., and Zhou, Z.-H.
\newblock Multi-layered gradient boosting decision trees.
\newblock In \emph{Neural Information Processing Systems}, 2018.

\bibitem[Fern{\'a}ndez-Delgado et~al.(2014)Fern{\'a}ndez-Delgado, Cernadas,
  Barro, and Amorim]{fernandez2014we}
Fern{\'a}ndez-Delgado, M., Cernadas, E., Barro, S., and Amorim, D.
\newblock Do we need hundreds of classifiers to solve real world classification
  problems?
\newblock \emph{Journal of Machine Learning Research}, 15\penalty0
  (1):\penalty0 3133--3181, 2014.

\bibitem[{Gonzalez-Guerrero} et~al.(2019){Gonzalez-Guerrero}, {Tracy}, {Guo},
  and {Stan}]{lowenergyrf}
{Gonzalez-Guerrero}, P., {Tracy}, T., {Guo}, X., and {Stan}, M.~R.
\newblock Towards low-power random forest using asynchronous computing with
  streams.
\newblock In \emph{International Green and Sustainable Computing Conference},
  pp.\  1--5, Oct 2019.
\newblock \doi{10.1109/IGSC48788.2019.8957193}.

\bibitem[Google(2020)]{googlearticle}
Google.
\newblock Minimizing real-time prediction serving latency in machine learning.
\newblock
  \url{https://cloud.google.com/solutions/machine-learning/minimizing-predictive-serving-\
  latency-in-machine-learning}, 2020.

\bibitem[Hellerstein et~al.(2018)Hellerstein, Faleiro, Gonzalez,
  Schleier-Smith, Sreekanti, Tumanov, and Wu]{hellerstein2018serverless}
Hellerstein, J.~M., Faleiro, J., Gonzalez, J.~E., Schleier-Smith, J.,
  Sreekanti, V., Tumanov, A., and Wu, C.
\newblock Serverless computing: One step forward, two steps back.
\newblock \emph{arXiv preprint arXiv:1812.03651}, 2018.

\bibitem[{Ishakian} et~al.(2018){Ishakian}, {Muthusamy}, and
  {Slominski}]{8360337}
{Ishakian}, V., {Muthusamy}, V., and {Slominski}, A.
\newblock Serving deep learning models in a serverless platform.
\newblock In \emph{2018 IEEE International Conference on Cloud Engineering
  (IC2E)}, pp.\  257--262, April 2018.
\newblock \doi{10.1109/IC2E.2018.00052}.

\bibitem[Jonas et~al.(2017)Jonas, Venkataraman, Stoica, and
  Recht]{DBLP:journals/corr/JonasVSR17}
Jonas, E., Venkataraman, S., Stoica, I., and Recht, B.
\newblock Occupy the cloud: Distributed computing for the 99{\%}.
\newblock \emph{CoRR}, abs/1702.04024, 2017.

\bibitem[Ke et~al.(2017)Ke, Meng, Finley, Wang, Chen, Ma, Ye, and
  Liu]{ke2017lightgbm}
Ke, G., Meng, Q., Finley, T., Wang, T., Chen, W., Ma, W., Ye, Q., and Liu,
  T.-Y.
\newblock {LightGBM}: A highly efficient gradient boosting decision tree.
\newblock In \emph{Advances in Neural Information Processing Systems}, pp.\
  3146--3154, 2017.

\bibitem[Kumar et~al.(2017)Kumar, Goyal, and Varma]{kumar2017resource}
Kumar, A., Goyal, S., and Varma, M.
\newblock Resource-efficient machine learning in {2 KB RAM for the Internet of
  Things}.
\newblock In \emph{International Conference on Machine Learning-Volume 70},
  pp.\  1935--1944. JMLR. org, 2017.

\bibitem[{Li} et~al.(2018){Li}, {Ota}, and {Dong}]{8270639}
{Li}, H., {Ota}, K., and {Dong}, M.
\newblock Learning {IoT} in edge: Deep learning for the {Internet of Things}
  with edge computing.
\newblock \emph{IEEE Network}, 32\penalty0 (1):\penalty0 96--101, 2018.
\newblock \doi{10.1109/MNET.2018.1700202}.

\bibitem[Mehta(2019)]{rapids}
Mehta, V.
\newblock Accelerating random forests up to 45x using cuml.
\newblock
  https://medium.com/rapids-ai/accelerating-random-forests-up-to-45x-using-cuml-dfb782a31bea,
  2019.

\bibitem[Mhembere et~al.(2017)Mhembere, Zheng, Priebe, Vogelstein, and
  Burns]{mhembere17knor}
Mhembere, D., Zheng, D., Priebe, C.~E., Vogelstein, J.~T., and Burns, R.
\newblock {knor}: A {NUMA}-optimized in-memory, distributed and
  semi-external-memory k-means library.
\newblock In \emph{High-Performance Parallel and Distributed Computing}, 2017.

\bibitem[Moritz et~al.(2018)Moritz, Nishihara, Wang, Tumanov, Liaw, Liang,
  Elibol, Yang, Paul, Jordan, et~al.]{moritz2018ray}
Moritz, P., Nishihara, R., Wang, S., Tumanov, A., Liaw, R., Liang, E., Elibol,
  M., Yang, Z., Paul, W., Jordan, M.~I., et~al.
\newblock Ray: A distributed framework for emerging {AI} applications.
\newblock In \emph{Symposium on Operating Systems Design and Implementation},
  2018.

\bibitem[Nakandala(2020)]{hummingbird}
Nakandala, Markus~Weimer, M.~I.
\newblock A tensor compiler approach for one-size-fits-all {ML} prediction
  serving.
\newblock In \emph{Symposium on Operating Systems Design and Implementation},
  2020.

\bibitem[Olston et~al.(2017)Olston, Fiedel, Gorovoy, Harmsen, Lao, Li,
  Rajashekhar, Ramesh, and Soyke]{olston2017tensorflow}
Olston, C., Fiedel, N., Gorovoy, K., Harmsen, J., Lao, L., Li, F., Rajashekhar,
  V., Ramesh, S., and Soyke, J.
\newblock Tensorflow-serving: Flexible, high-performance {ML} serving.
\newblock \emph{arXiv preprint arXiv:1712.06139}, 2017.

\bibitem[Painsky \& Rosset(2018)Painsky and Rosset]{painsky2018lossless}
Painsky, A. and Rosset, S.
\newblock Lossless (and lossy) compression of random forests.
\newblock \emph{arXiv preprint arXiv:1810.11197}, 2018.

\bibitem[Pedregosa et~al.(2011)Pedregosa, Varoquaux, Gramfort, Michel, Thirion,
  Grisel, Blondel, Prettenhofer, Weiss, Dubourg, et~al.]{pedregosa2011scikit}
Pedregosa, F., Varoquaux, G., Gramfort, A., Michel, V., Thirion, B., Grisel,
  O., Blondel, M., Prettenhofer, P., Weiss, R., Dubourg, V., et~al.
\newblock Scikit-learn: Machine learning in python.
\newblock \emph{Journal of Machine Learning Research}, 12\penalty0
  (Oct):\penalty0 2825--2830, 2011.

\bibitem[Ribeiro et~al.(2015)Ribeiro, Grolinger, and Capretz]{inproceedings}
Ribeiro, M., Grolinger, K., and Capretz, M.
\newblock Mlaas: Machine learning as a service.
\newblock 12 2015.
\newblock \doi{10.1109/ICMLA.2015.152}.

\bibitem[Stasior et~al.(2017)Stasior, Carson, Dasari, and Kim]{stasior2017zero}
Stasior, W.~F., Carson, D.~A., Dasari, R., and Kim, Y.
\newblock Zero latency digital assistant, March~9 2017.
\newblock US Patent App. 15/147,726.

\bibitem[Symanovich(2020)]{iotfuture}
Symanovich, S.
\newblock The future of iot: 10 predictions about the internet of things.
\newblock
  https://us.norton.com/internetsecurity-iot-5-predictions-for-the-future-of-iot.html,
  2020.

\bibitem[Tzimpragos et~al.(2019)Tzimpragos, Madhavan, Vasudevan, Strukov, and
  Sherwood]{tzimpragos2019boosted}
Tzimpragos, G., Madhavan, A., Vasudevan, D., Strukov, D., and Sherwood, T.
\newblock Boosted race trees for low energy classification.
\newblock In \emph{International Conference on Architectural Support for
  Programming Languages and Operating Systems}, pp.\  215--228, 2019.

\bibitem[Zhao et~al.(2019)Zhao, Deng, Zhang, and Yang]{zhao2019rfacc}
Zhao, L., Deng, Q., Zhang, Y., and Yang, J.
\newblock {RFAcc: A 3D ReRAM} associative array based random forest
  accelerator.
\newblock In \emph{International Conference on Supercomputing}, pp.\  473--483,
  2019.

\bibitem[Zhou \& Feng(2017)Zhou and Feng]{zhou17deepforest}
Zhou, Z.-H. and Feng, J.
\newblock Deep forest: Towards an alternative to deep neural networks.
\newblock In \emph{Proceedings of the International Joint Conference on
  Artificial Intelligence}, 2017.

\end{thebibliography}
\bibliographystyle{mlsys2020}

%MMTODO:
%3)Figure 2: blockwise layout needs to be edited
%4)Fix bibliography warnings
%appendix for layout algorithms
%make figures smaller
%Page limit

%%%%%%%%%%%%%%%%%%%%%%%%%%%%%%%%%%%%%%%%%%%%%%%%%%%%%%%%%%%%%%%%%%%%%%%%%%%%%%%
%%%%%%%%%%%%%%%%%%%%%%%%%%%%%%%%%%%%%%%%%%%%%%%%%%%%%%%%%%%%%%%%%%%%%%%%%%%%%%%
% SUPPLEMENTAL CONTENT AS APPENDIX AFTER REFERENCES
%%%%%%%%%%%%%%%%%%%%%%%%%%%%%%%%%%%%%%%%%%%%%%%%%%%%%%%%%%%%%%%%%%%%%%%%%%%%%%%
%%%%%%%%%%%%%%%%%%%%%%%%%%%%%%%%%%%%%%%%%%%%%%%%%%%%%%%%%%%%%%%%%%%%%%%%%%%%%%%
\pagebreak
\appendix
\section{Appendix}

Here we describe additional experiments in detail for PACSET as a service. The two experiments are determining the ideal Redis block size and to investigate the effects of running multiple concurrent lambda jobs.

        \begin{figure}[h]
                \includegraphics[width=\columnwidth]{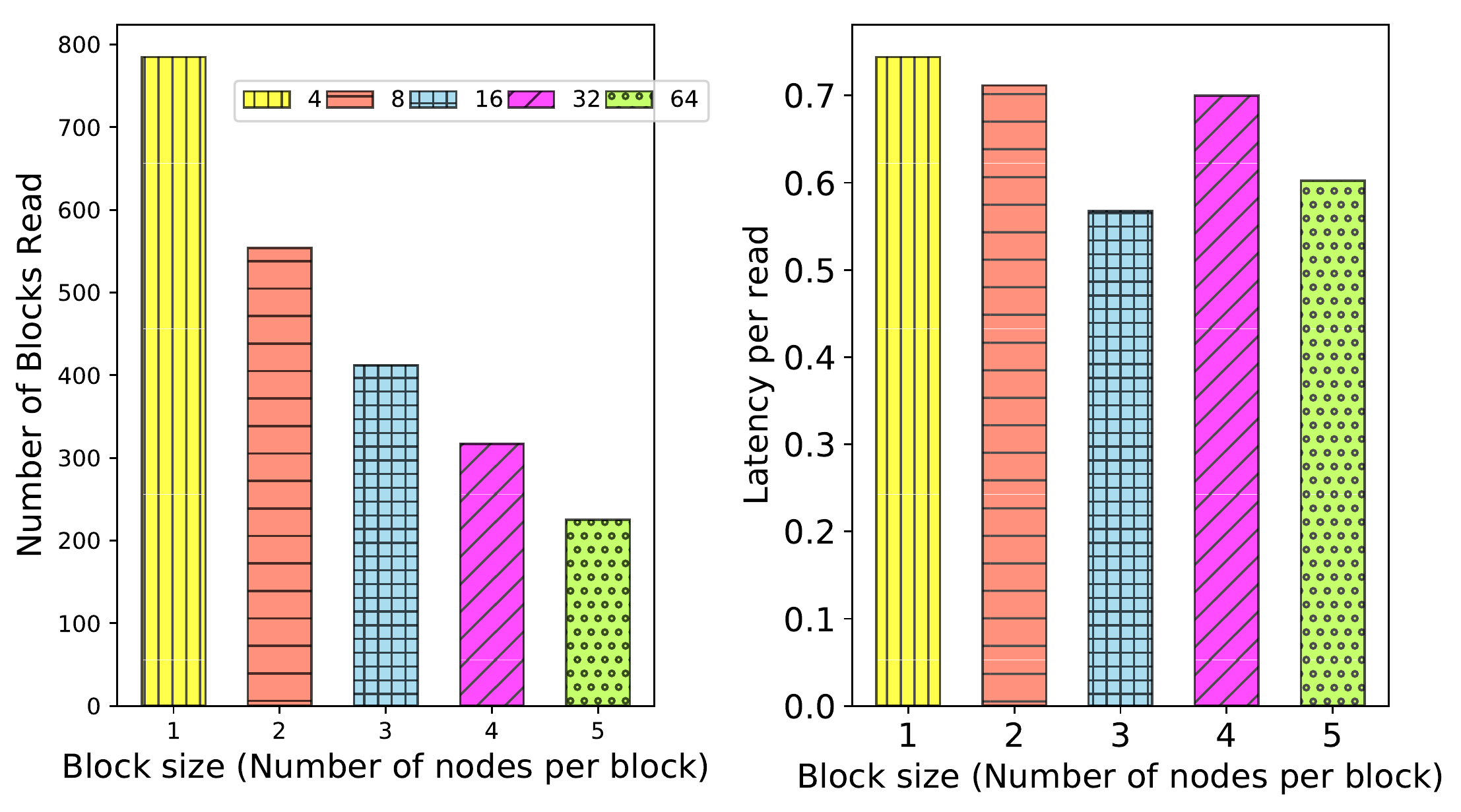}
                
                    \vspace{-8pt}
                    
                \caption{Number of blocks and inference per block as a function of block width}
                \label{fig:lambdablock}
         \end{figure}

            \begin{figure}[h]
            \includegraphics[width=\columnwidth]{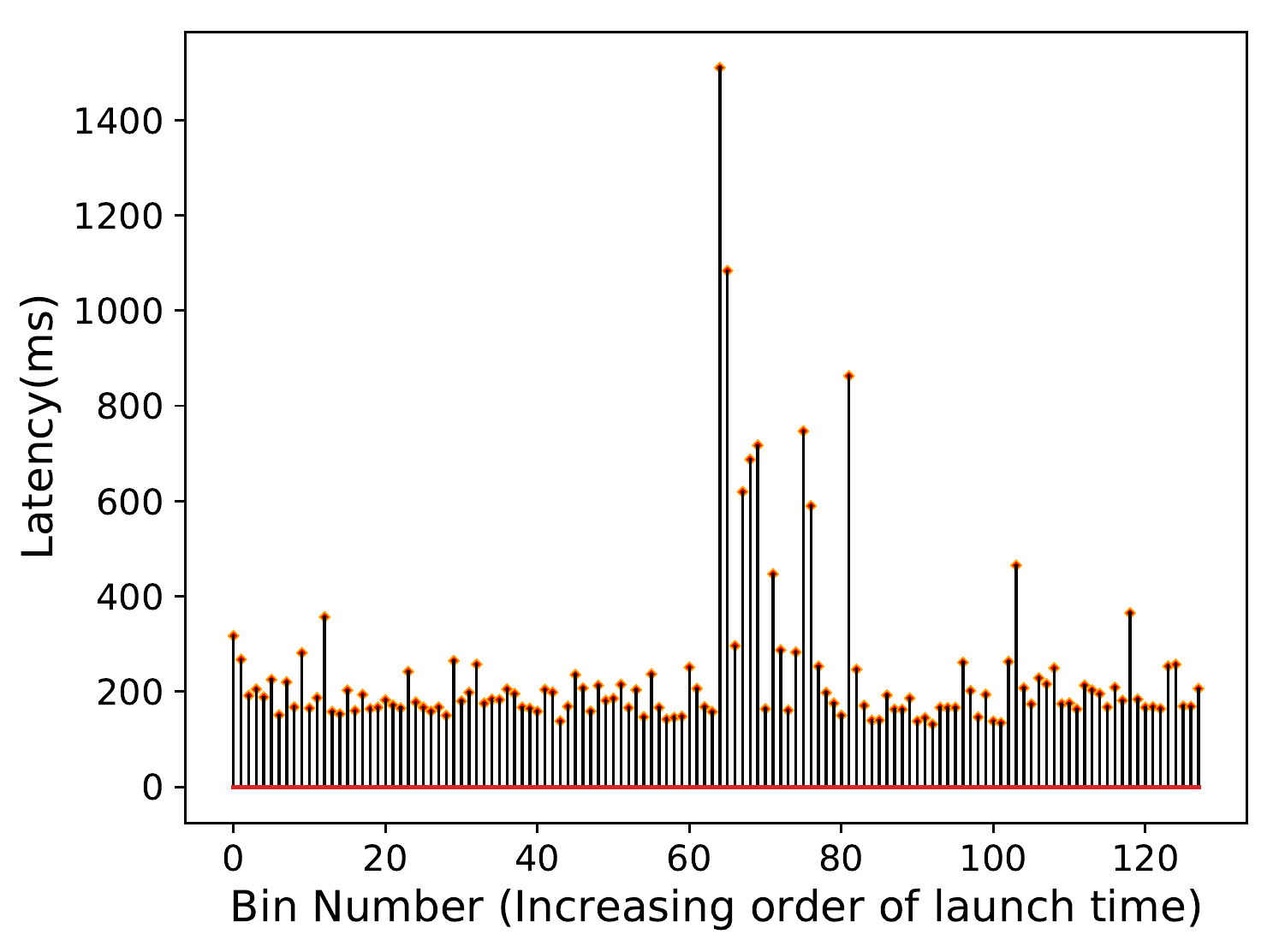}
            
            \caption{Timeline of Bin inference}\label{fig:timeline}
        \end{figure}

                \def\names{
        {Lambda_total_bar},
        {Lambda_128_hist}}
    
        \begin{figure}[t!]
            \foreach \name in \names {%
                \begin{subfigure}[p]{\columnwidth}
                    \includegraphics[width=\columnwidth]{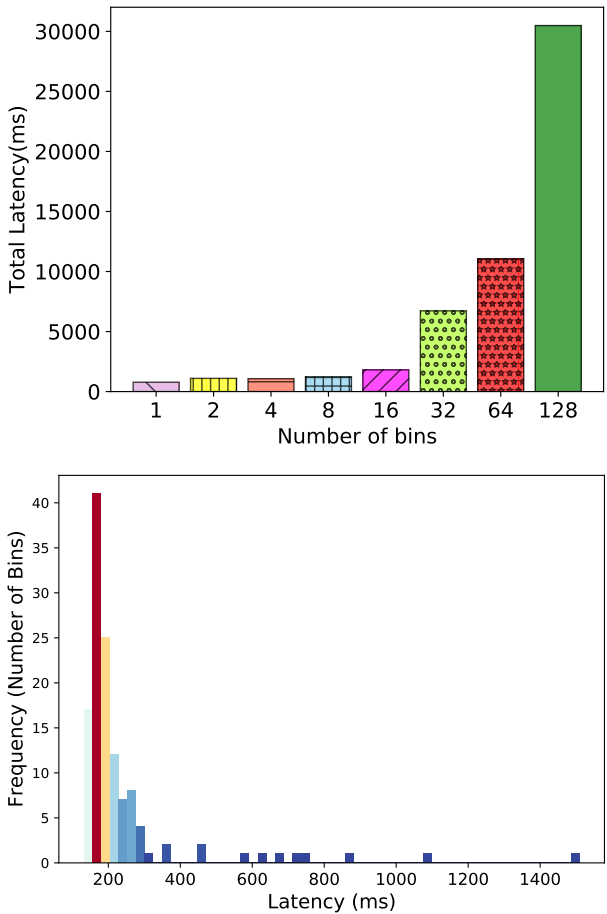}
                    
                \end{subfigure}
            }
            \caption{Histogram of bin inference latencies}\label{fig:lambdaconc}
            
        \end{figure}

    \subsubsection{Ideal Redis Block size (Size of key, value pair/ num nodes in a key-value pair)}\label{exp:pacsetaasblock}
       
Here, we would like to investigate the effect of the number of nodes stored per key (bucket size) on the latency. We vary the bucket size and compute cold start inference latency as a function of bucket size. Figure~\ref{fig:lambdablock}
shows the number of blocks read (Left) and the latency per read (Right) as a function of the block size. Here, a single read corresponds to reading one key value pair. We had initially expected the ideal blocksize to be large because then we can save I/O by paying up front. On the contrary, the results show that smaller blocksizes are preferred, more specifically,  around 16 seem to be ideal.  This can be rationalized as follows. Since we are performing fine-grained I/O, i.e the number of nodes which we will actually access during inference is only a small fraction of the total number of nodes, smaller block sizes work better. Small block sizes along with blockwise packing together ensure that each block read contains as much useful information as possible. However, if the block size is too small, then we end up making many calls to Redis and incur a high I/O cost. This is a trade-off that needs to be made when choosing the ideal block size.  The layout used is the BIN+blockwise layout.

    \subsubsection{Lambda concurrency}\label{lambda:conc}

In this experiment, we investigate the effect of deploying multiple lambda functions concurrently, each on a subset of trees. Here, we would like measure the speedup achieved by parallelizing the inference across bins . We invoke the lambda functions asynchronously and concurrently such that each invocation performs inference on a bin (subset of trees) in parallel. Tree ensemble inference is embarrassingly parallel. Thus,  we had initially expected there to be almost linear speed-up. On the contrary, there were quite a few factors against parallelism that that we saw. First, there was skew as a result of AWS lambda function scheduling of which the user has no control. This resulted in the invocations not all being scheduled at the same time. The last scheduled job and the first scheduled job are scheduled seconds apart. Since they perform the same amount of work this results in a decrease in true parallelism. Some of the initial latency spikes can be attributed to cold start latencies. Figure ~\ref{fig:lambdaconc} on the left shows the total inference latency for various degrees of concurrency.  Figure ~\ref{fig:lambdaconc} on the right shows a histogram of bin latencies with 128 trees and 128 concurrent lambda functions. We see a large latency spike in the middle which is where the most throttling occurs due to delays in scheduling. Figure ~\ref{fig:timeline} shows the time taken for inference of each bin numbered by the lambda invocation order corresponding to the bin. We see that the bins in the middle take the longest because around this time, all 128 lambda invocations are running together and accessing Redis simultaneously leading to contention and thus slowing down the inference time.

%%%%%%%%%%%%%%%%%%%%%%%%%%%%%%%%%%%%%%%%%%%%%%%%%%%%%%%%%%%%%%%%%%%%%%%%%%%%%%%
%%%%%%%%%%%%%%%%%%%%%%%%%%%%%%%%%%%%%%%%%%%%%%%%%%%%%%%%%%%%%%%%%%%%%%%%%%%%%%%

\end{document}